\documentclass{emulateapj}

\shorttitle{SDSSJ095209.56+214313.3}
\shortauthors{Komossa et al.}

\begin{document}

\title{$NTT$, $Spitzer$ and $Chandra$ spectroscopy of
       SDSSJ095209.56+214313.3: 
       the most luminous coronal-line supernova ever observed, or a stellar tidal disruption event ?}

\author{S. Komossa\altaffilmark{1}, H. Zhou\altaffilmark{1,6}, A. Rau\altaffilmark{2}, M. Dopita\altaffilmark{3}, 
A. Gal-Yam\altaffilmark{4}, J. Greiner\altaffilmark{1}, J. Zuther\altaffilmark{5}, M. Salvato\altaffilmark{2}, 
D. Xu\altaffilmark{7}, H. Lu\altaffilmark{1,6},
R. Saxton\altaffilmark{8}, M. Ajello\altaffilmark{9,10}}  

\altaffiltext{1}{Max-Planck-Institut f\"ur extraterrestrische Physik, Postfach 1312, 85741 Garching, Germany; skomossa@mpe.mpg.de}
\altaffiltext{2}{California Institute of Technology, 105-24 Robinson 1200 E. California Blvd., Pasadena, CA 91125}
\altaffiltext{3}{Research School of Astronomy and Astrophysics, The Australian 
National University, Cotter Road, Weston Creek, ACT 2611, Australia} 
\altaffiltext{4}{Benoziyo Center for Astrophysics, Weizmann Institute of Science, 76100 Rehovot, Israel} 
\altaffiltext{5}{I. Physikalisches Institut, Universit\"at zu K\"oln, Z\"ulpicher Str. 77, 50937 K\"oln, Germany}
\altaffiltext{6}{Center for Astrophysics, University of Science and
 Technology of China, Hefei, Anhui, 230026, China}
\altaffiltext{7}{National Astronomical Observatories, Chinese Academy of Science,
A20 Datun Road, Chaoyang District, Beijing 100012, China}
\altaffiltext{8}{ESA/ESAC, Apartado 78, 28691 Villanueva de la Canada, Madrid, Spain}
\altaffiltext{9}{SLAC National Accelerator Laboratory, 2575 Sand Hill Road, Menlo Park, CA 94025}
\altaffiltext{10}{KIPAC, 2575 Sand Hill Road, Menlo Park, CA 94025}

\begin{abstract}
The galaxy 
SDSSJ095209.56+214313.3 (SDSSJ0952+2143 hereafter)
showed remarkable emission-line and continuum properties
and strong emission-line variability first reported in 2008 (paper I).
The spectral properties and low-energy variability
are the consequence of
a powerful high-energy flare which was itself not observed directly.
Here we report follow-up optical, near-infrared, mid-infrared, and 
X-ray observations of SDSSJ0952+2143.
We 
discuss  outburst scenarios in terms of stellar tidal disruption
by a supermassive black hole, peculiar variability
of an AGN, 
and a supernova explosion, and possible links between these scenarios and mechanisms.
The optical
spectrum of SDSSJ0952+2143 exhibits several peculiarities: an exceptionally high ratio of 
[FeVII] transitions over [OIII], a dramatic decrease by a factor of 10 of 
the highest-ionization coronal lines, a very unusual and variable
Balmer line profile including a triple-peaked 
narrow component with two unresolved horns,
and a large Balmer decrement.   
The MIR emission measured with the $Spitzer$ IRS in the narrow 10--20$\mu$m band
is extraordinarily luminous and amounts to $L_{\rm 10-20\mu m} = 3.5 \times 10^{43}$ erg\,s$^{-1}$.
The IRS spectrum shows a bump around $\sim$11$\mu$m and an increase toward longer wavelengths,
reminiscent of silicate emission.
The strong MIR excess over the NIR implies the dominance of  
relatively cold dust. The pre- and post-flare NIR host galaxy colours
indicate a non-active galaxy.  
The X-ray luminosity of $L_{\rm x,0.1-10keV} = 10^{41}$ erg\,s$^{-1}$ measured with $Chandra$ 
is below that typically observed in active galactic nuclei (AGN).  
Similarities of SDSSJ0952+2143 with some extreme supernovae (SNe) 
suggest the explosion of a supernova of Type IIn.
However, an extreme accretion event in a low-luminosity AGN or inactive galaxy,
especially stellar tidal disruption, remain possibilities, which 
could potentially produce a very similar
emission-line response.  
If indeed a supernova, SDSSJ0952+2143 is one of the most distant
X-ray and MIR detected SNe known so far, the most MIR luminous, and
one of the most X-ray luminous.  
It is also by far the most luminous ($>10^{40}$ erg\,s$^{-1}$)
in high-ionization coronal lines,   
exceeding previous SNe by at least a factor of 100.  
\end{abstract}

\keywords{circumstellar matter -- supernovae: general -- galaxies: 
emission lines -- galaxies: individual (SDSSJ095209.56+214313.3)}  

\section{Introduction}

Transient phenomena 
have been detected in various astrophysical object classes.
They are especially pronounced in the high-energy regime,
and provide us with important information on the physics of
astrophysical sources under extreme conditions.
Many of the transient phenomena are linked, in one way or another, to
the presence of compact objects and especially black holes and the physical
processes in their immediate environment.   
Transience in the high-energy domain covers a broad range of timescales
ranging from typically milli-seconds -- seconds 
in  Gamma-Ray Bursts (GRBs; Klebesadel et al. 1973, Piran 2005), 
over minutes -- hours in X-ray bursts (Grindlay et al. 1976, Lamb 2000) and early 
stages of supernova explosions (Soderberg et al. 2008), 
to weeks -- months in stellar tidal disruptions (Rees 1988, Komossa \& Bade 1999),
for instance. 
Other transient events have been predicted by theory, but not
yet observed, including long-lasting accretion disk flares of
recoiling supermassive black holes (SMBHs; Shields \& Bonning 2008), or very short-timescale
hard X-ray flares due to tidal detonations of stars (Brassart \& Luminet 2008).  

Powerful outbursts of radiation do not only carry key information about their
production mechanisms, they also can be used as a probe of their
gaseous environment: the radiation is reprocessed into ultraviolet (UV), optical,
infrared (IR) and X-ray emission lines which contain a wealth of information on the gaseous kinematics,
chemical composition, density and geometry of the line-emitting gas.    
In the context of supernova explosions, reprocessing of
high-energy radiation into emission lines carries information on  
the progenitor star and the circumstellar medium (Chugai \& Danziger 1994; Filippenko 1997).
Accretion flares from stars tidally disrupted by massive black holes
will illuminate the stellar post-disruption debris and the interstellar medium,
potentially including broad and narrow-line regions and molecular tori at the
cores of galaxies. 

We have found SDSSJ0952+2143 (Komossa et al. 2008) at redshift $z$=0.079
in a systematic search for emission-line galaxies    
in SDSS-DR6 (Sloan Digital Sky Survey Data Release 6; Adelman-McCarthy et al. 2008).  
The optical spectrum of SDSSJ0952+2143 and its multi-wavelength properties 
turned out to be exceptional.
The optical spectrum is dominated by strong iron coronal lines
with the highest ratios of several [FeVII] transitions over 
[OIII]5007 measured among galaxies. 
The H$\alpha$ profile shows a multi-peaked structure.
The highest-ionization iron lines have significantly faded between
2005 and 2007, and the optical continuum emission has decreased.  
These properties were interpreted as an emission-line and continuum
response to a powerful high-energy flare (Komossa et al. 2008; hereafter paper I)
which itself escaped detection. 
After the initial recognition of the unusual spectrum and
quick optical follow-up spectroscopy in December 2007,
we have initiated a number of multi-wavelength observations,
especially having in mind the possibility of using the emission-line
response to the flare as a rare chance to do reverberation mapping
of the different systems of gaseous matter in the galaxy core. 
Here, we report follow-up spectroscopy in the IR, optical and X-ray band,
and explore outburst scenarios. The paper is structured as
follows: In Sect. 2  we present the new optical, IR and X-ray observations 
and describe the data analysis. In Sect. 3 we use 
emission-line and continuum
properties as diagnostics of the physical conditions in the line-emitting gas. 
Sect. 4 scrutinizes scenarios which could explain
the unusual multi-wavelength properties of SDSSJ0952+2143, with focus
on a supernova explosion, exceptional AGN-type variability and tidal-disruption-related scenarios. 
The conclusions
are provided in Sect. 5. 
In Tab. 1 an overview of the previous and new observations
and the sequence of events is presented. 
We use a cosmology (Wright 2006) with
$H_{\rm 0}$=70 km\,s$^{-1}$\,Mpc$^{-1}$, $\Omega_{\rm M}$=0.3
and $\Omega_{\rm \Lambda}$=0.7 throughout this paper.

\section{Multi-wavelength observations}

\subsection{New Technology Telescope ($NTT$) optical spectroscopy}

\subsubsection{Data reduction}

We have observed SDSSJ0952+2143 with $EMMI${\footnote{see the EMMI user's manual
at  
http://www.ls.eso.org/docs/\\LSO-MAN-ESO-40100-0001/LSO-MAN-ESO-40100-0001.pdf}} at the 
ESO-$NTT$ 3.5m telescope on 2008 February 8  (hereafter referred to as 2008 $NTT$ spectrum). 
We took two exposures each
of 1500 s duration with grism
 \#5 (wavelength range: 3800\AA -- 7000\AA) and \#6
(wavelength range: 5800\AA -- 8600\AA), 
respectively,
using a slit width of 1.5 arcsec oriented at the parallactic angle 
(Filippenko 1982). 
This setting gave a spectral
resolution of $\sim$6\AA.
KPNO standard stars were observed before and after each target exposure
for flux calibration.
He-Ne-Ar lamp spectra  were obtained with the
two grisms in order to carry out the wavelength calibration.   

The raw two-dimensional data were reduced with standard
procedures using the software package IRAF{\footnote{IRAF is distributed by the 
National Optical Astronomy Observatory, which
is operated
by the Association of Universities for Research in Astronomy, Inc., under
cooperative
agreement with the National Science Foundation.}}. 
The CCD data reduction includes bias subtraction, flat-field correction, 
and cosmic-ray removal. 
The task {\it apall}  was used to extract the spectra. 
We then carried out the wavelength and flux calibration using the 
He-Ne-Ar lamp spectra and the standard stars. 
After flux calibration the two spectra
obtained with grism\#5 agree within 4\% with each other, and with
one of the two grism\#6 spectra within 8\%. The second grism\#6 exposure
deviated in its normalization which could be traced back to a change in seeing during
that target exposure, and this spectrum was therefore shifted in flux scale to match the
other 3 spectra.  
The spectra were corrected for  Galactic extinction of E(B-V)=0.028.
All four spectra were then combined and
the resulting spectrum is shown in Fig. 1 where it is compared 
with the previous SDSS spectrum which was taken in 2005. 
Telluric absorption was corrected for, using the standard stars and 
taking into account the airmass during the observations.

\subsubsection{Underlying spectral energy distribution (SED)}

Using the flux calibration scheme described above, the observed continuum level, 
dominated by starlight from the host galaxy, is lower 
in the $NTT$ spectrum than in the SDSS spectrum, which can be traced back
to aperture loss due to the extended host galaxy.
The flux in the [OIII]5007 emission line is consistent with being constant
between the 2005 SDSS and 2008 ESO $NTT$ observation (Tab. 1). 
In any case, when using emission-line ratios
for diagnostics, we preferentially use neighbouring
emission lines, to be more independent of residual absolute calibration
uncertainties.

The SDSS and $NTT$ spectra were decomposed into a stellar and a non-stellar
(powerlaw) component using the method described  
by Lu et al. (2006, see also Zhou et al. 2006), 
and are both dominated by the stellar component
of the host galaxy. 
The continuum emission from the host galaxy 
can be well fit with a single stellar
population with an age of $T \simeq 2$ Gyr.
The underlying non-stellar continuum component is very faint in the $NTT$ spectrum.
In paper I, we used instead a  mix of synthesized galaxy template spectra
to describe the host galaxy emission, rather than one single stellar population.
Re-modelling of the continuum introduces slight changes (typically $<$10--20\%) in the re-measured
emission lines reported here. 

\subsubsection{Narrow emission lines} 

After the SED was decomposed into host galaxy and powerlaw contribution (Fig. 1),
these components were subtracted from the spectrum and the emission lines were then measured. 
Emission-line widths reported here and in paper I have been corrected for instrumental
broadening. 
Most emission lines which were present in the 2005 SDSS spectrum are
still detected in the 2008 $NTT$ spectrum, even though a number of
them, especially the highest-ionization lines, the broad Balmer lines,  
and the peculiar narrow Balmer horns are significantly fainter (see below).
We identify transitions of [OI], [OII], [OIII], [NII], [SII], HeII, 
[NeIII], [FeVI], Fe[VII], [FeX], [FeXI], [FeXIV], and Balmer lines (Figures 1-4).  

We have first refit all emission lines of the 2005 SDSS spectrum
with a single Gaussian component, and then did the same for the 2008 $NTT$ spectrum.  
Single Gaussian fits to those emission lines which do not
show spectral complexity  
result in line widths
of 200-300 km\,s$^{-1}$ 
in the SDSS spectrum. 
These same lines are not resolved in the $NTT$  
spectrum, which was of lower resolution, consistent
with constant line width. Most of the high-ionization
lines appear broader than the low-ionization lines, 
indicative of a two-component nature
of their line profiles (see below). FWHMs based on single Gaussian fits listed in Tab. 2
should therefore not be used per se, but should rather be taken as an indication of
a more complex emission-line profile, if the line width exceeds $\sim$300 km\,s$^{-1}$.  

As already noted in paper I, the highest-ionization iron lines have strongly
faded. Here we greatly improve limits on [Fe X] and [Fe XIV] and for the first time
include [FeXI] in the $NTT$ spectral band, in comparison with the earlier 2007 $Xinglong$
follow-up spectroscopy. 
[FeXI] is a factor $\sim$10 fainter in the 2008 $NTT$ spectrum than it was in
the 2005 SDSS spectrum. [FeX] is
very faint, even though the exact measurement is
hampered by uncertainties in the precise correction of the superposed 
telluric (atmospheric) absorption. 

The line emission from 
[OIII]4363 and HeII4686 is 
remarkably strong in the 2005 and 2008 spectra, with intensity
ratios [OIII]4363/[OIII]5007 = 0.2, and HeII4686$_{\rm totl}$/H$\beta_{\rm n}$ = 2 in 2005
(while HeII4686$_{\rm n}$/H$\beta_{\rm n}$ = 0.3, below the value of 0.66 above which the ratio
is dominated by helium abundance; Dopita \& Sutherland 2003). Here, the index 
{\em totl} refers to the total line emission, while the index {\em n}
refers to the emission of the narrow component.  
HeII4686$_{\rm totl}$/H$\beta_{\rm n}$ decreased by a factor 2.8 which can be traced back
to a decrease in the broad component in HeII (Sect. 2.1.4).

We have compared the 2007 $Xinglong$ spectrum (paper I) with the 2008 $NTT$ spectrum in
order to search for spectral variations on the timescale of two months. Keeping
in mind the lower S/N and resolution of the $Xinglong$ spectrum, emission lines are
consistent with being constant within better than a factor of 2. 
A new optical spectrum was acquired with the 2.16m $Xinglong$ telescope
on 2008 December 25 with an exposure time of 7200 s. The strongest emission lines
are still all present and no new spectral features have emerged. 
[FeVII]5722 is no longer safely detected, indicating a decrease of this line
at the 3$\sigma$ level. 
The full results of
this and future optical monitoring will be presented elsewhere.   

\subsubsection{Two-component high-ionization lines}
When fit by single Gaussians, several of the high-ionization lines
show a relatively  broad profile, and are in fact better modeled by
two components; a broad base and a narrow core. Seven lines in the SDSS spectrum
are strong enough for such a decomposition; while for only two of them the decomposition
can still be done with the $NTT$ spectrum.  These seven lines are [FeVII]3759, HeII4686, [OIII]5007,
[FeVII]5722, [FeVII]6087, [FeX]6376 and [FeXI]7894. If the narrow core of each line is fixed to the FWHM 
of [SII], 210 km\,s$^{-1}$, the broad base is well fit by a second 
Gaussian with a width of $\sim$600--800 km\,s$^{-1}$ ([FeVII]5722, [FeX]6376, [FeVII]3759 and [FeXI]7894),
900--1100 km\,s$^{-1}$ ([FeVII]6087 and HeII), 
and 400 km\,s$^{-1}$ ([OIII]5007), respectively. 
[FeVII]6087 and HeII4686 can still be decomposed in
the ESO spectrum and were represented by a two-component Gaussian 
assuming that neither the width of the broad component nor that of the narrow component
varied between 2005 and 2008 
(and taking into account the different instrumental broadening). 
We find that the flux in the narrow component 
is almost constant, while the broad component decreased by at 
least a factor of $\sim$3 (Fig. 4).

\subsubsection{Balmer line profile and variability}

We paid special attention to changes in the very unusual profile
of the Balmer lines, which consist of several components;
an asymmetric broad base, a narrow core, and two strong peculiar unresolved horns. 
We carefully decomposed the line profiles, 
and searched for changes between the 2008 $NTT$ spectrum and the 2005 SDSS spectrum.
The H$\alpha$+[NII] blend in the SDSS spectrum was fit with seven Gaussians:
(1) Three Gaussians for the ``normal'' narrow emission lines in H$\alpha$ and [NII]6548,6584.
The redshift and line widths were fixed to that of the [SII]6716,6731
doublet, and the [NII]6548/6584 ratio was fixed to the theoretical
value of 1/3.  
(2) Two Gaussians for the narrow horns, with line width fixed to the
instrumental
resolution, and centroids and intensities as free parameters{\footnote{In what follows
we will generally assume that only these two strong horns are present, but note here
that we cannot exclude the presence of more such components at very
faint emission levels.}}. 
(3) Two Gaussians for the broad component of H$\alpha$ with centroids,
widths, and intensities as free parameters.  
Since the broad base is much broader than the narrow components
and since the narrow horns are red- and blueshifted and do not
coincide with
the narrow core of H$\alpha$, 
the broad and narrow components can
be relatively well decomposed. Errors in FWHM
are typically less than 20\%. At the same time, the two
Gaussian components which make up the broad base of
H$\alpha$ are not well constrained individually. We therefore
only report the FWHM of the sum of the two Gaussians
that fit this component, and only report the total
line flux, but no results on the two components
separately.   

The H$\beta$ regime was fit with five Gaussians, that match the components in H$\alpha$.
(1) One Gaussian for the normal narrow H$\beta$  emission line, with its 
centroid redshift and width
fixed to that of the corresponding H$\alpha$ line. 
(2) Two Gaussians for the narrow horns, also with their centroids and widths
fixed to that of H$\alpha$.
(3) Two Gaussians for the broad component of H$\beta$, with their centroids, widths,
and the intensity ratio of the two Gaussians fixed to that of the broad component of H$\alpha$. 
The results of the decomposition are shown in Fig. 3.

The H$\alpha$ and H$\beta$ complexes in the 2008 $NTT$ spectrum were fit using the same
scheme as for the 2005 SDSS spectrum, but all the widths of the narrow lines (including
the narrow horns) which are unresolved, were fixed
to the instrumental broadening. The two Gaussians for the broad component
were not fixed to the SDSS values, and indeed both the intensity and profile
of H$\alpha$ changed significantly between the two observations, with an
intensity ratio of 0.35 between the ESO and SDSS observation. 
The broad H$\alpha$ component is
still asymmetric with excess emission in the red
part of the line
with a centroid shift relative to the narrow core
of $v_{\rm SDSS} \simeq 560$ km\,s$^{-1}$ (SDSS spectrum) 
in comparison to $v_{\rm NTT} \simeq 270$ km\,s$^{-1}$ ($NTT$ spectrum), while the line width
changed from FWHM $\simeq 2100$ km\,s$^{-1}$  (SDSS) to FWHM $\simeq$ 1500 km\,s$^{-1}$ ($NTT$). 
The broad component in H$\beta$ is no longer safely detected in the
$NTT$ spectrum.

The remarkable narrow horns remained fixed in red/blueshift
(at $v$ = 540 and $-340$ km\,s$^{-1}$, where a negative sign indicates blueshift), 
but got fainter
(Tab. 2). 

H$\gamma$ is too faint for any decomposition,
and we merely report results from a single Gaussian fit in Tab. 2. 

\subsection{$GROND$ JHK photometry} 

Photometry was performed with the 7-channel imager $GROND$ (Greiner et al. 2008) 
attached to the 2.2m telescope
at La Silla in the filters g, r, i, z and J, H and K$_{\rm s}$.
The observations were carried out 
on 2008 January 1, at a seeing of 1$^{\prime\prime}$.
Each observation consisted 
of 4 exposures with 46 s each in the visual and 24 exposures with
10 s each in the NIR channels, giving an effective exposure time
of 4 min in the NIR and 3 min in the visual.
The images were flatfield and bias corrected using standard IRAF
routines. While the images in the visual bands are simply stacked
after astrometric registration, the NIR images are first distortion
corrected and then co-added with the shift-and-add method of the
jitter command in the eclipse package (Devillard 1997).   

Results from g,r,i, and z were reported in paper I which
showed that the luminosity of the galaxy core decreased by a factor 1.5-2 between the
2004 SDSS and the 2008 $GROND$ photometry. Here we add the NIR results.
The J, H and K$_{\rm s}$ photometry was calibrated against the 2MASS 
measurement of 1998. 
Given the different spatial resolutions of $GROND$ and 2MASS, only the integrated emission
can be compared. 
The total J,H and K$_{\rm s}$ entries listed in the 2MASS catalogue
of extended objects 
and cited in NED (Skrutskie et al. 2006) are J,H,K$_{\rm s}$ = (15.4,14.5,14.4) mag.  
We have selected sources in the field of view
which have comparable brightness as SDSSJ0952+2143,
and have used these to perform the $GROND$ photometry.
Given the faintness of the galaxy in the 2MASS exposure,
we conservatively assign a photometric error of 0.5 mag.
Within this error, the $GROND$ magnitudes are consistent with
2MASS. 

The J (H) coordinates of the galaxy centre (J2000), RA = 09:52:09.54 (09:52:09.56) 
and DEC = +21:43:13.2 (+21:43:13.3), are 
consistent with the SDSS position.  Radial profiles of the host galaxy
of SDSSJ0952+2143 in comparison with the PSF of nearby stars show that the 
emission from the galaxy can be traced out to a radius of 3--4$^{\prime\prime}$ in the J band.  
Close inspection of the J,H,K$_{\rm s}$ images reveals the emergence of a spiral structure at faint 
emission levels
in all three images (Fig. 5).  
Finally, we note that no other bright NIR source 
within a few arcseconds of SDSSJ0952+2143, which could
have affected the 2MASS measurements,
is detected (Fig. 5).

\subsection{$Chandra$, $XMM-Newton$ and $ROSAT$ X-ray observations}  

SDSSJ0952+2143 was observed and not detected during the $ROSAT$ all-sky survey  
in November 1990.  The upper limit on its PSPC countrate, $<$0.036 cts\,s$^{-1}$,
translates into an upper limit on its (0.1-2.4) keV X-ray luminosity of
$L_{\rm x} < 10^{43}$ erg\,s$^{-1}$ (assuming an X-ray powerlaw with photon
index $\Gamma_{\rm x}=-1.9$ and no excess absorption above the Galactic value, 
$N_{\rm H,gal} = 2.79 \times 10^{20}$ cm$^{-2}$).   

The field of SDSSJ0952+2143 was also covered during an $XMM-Newton$ slew observation, slew 9044100004,
on 2002 May 7.  No photons from the source were detected, with an EPIC pn upper limit  
of 1.3 cts\,s$^{-1}$. 
Again assuming a spectrum with $\Gamma_{\rm x}=-1.9$ and $N_{\rm H,gal}$ 
gives an upper limit on the (0.2-10 keV) X-ray luminosity of $L_{\rm x} < 8 \times 10^{43}$ erg\,s$^{-1}$.

SDSSJ0952+2143 was also not detected in X-rays 
during the $Swift$ BAT survey (Markwardt et al. 2005, Ajello et al. 2008).
The  
upper limit from 3 years of BAT survey observations between March 2005 and March 2008
is $f_{\rm x} < 8 \times 10^{-12}$ erg\,cm$^{-2}$\,s$^{-1}$  
in the 15-55 keV band, or $L_{\rm x} < 1 \times 10^{44}$ erg\,s$^{-1}$ in the same band. 

After the $Xinglong$ and $GROND$ detection of variability, we initiated
a $Chandra$ DDT observation (paper I; ObsId 9814) which was carried out
quasi-simultaneous with the $NTT$ observation (Tab. 1). 
Had the source still been sufficiently bright in X-rays, this would have provided
a unique chance to ``reverberation-map'' responses of the high-ionization lines to changes
in the ionizing continuum. 
During the 10 ks ACIS-S observation, 
only faint X-ray emission was detected with a countrate of 7\,10$^{-4}$ cts\,s$^{-1}$.
Among the detected photons,  
$\sim$40\% were above 3 keV; indicating a relatively
hard X-ray spectrum, which is not strongly absorbed.
Since the source is too faint to perform spectral fitting,
we have used two characteristic models in order to estimate fluxes
and luminosities.  
Assuming a spectrum with $\Gamma_{\rm x}=-1.9$ and $N_{\rm H,gal}$
gives an X-ray luminosity of $L_{\rm x} = 4 \times 10^{40}$ erg\,s$^{-1}$ (2-10 keV)
and $L_{\rm x} = 1 \times 10^{41}$ erg\,s$^{-1}$ (0.1-10 keV). Using instead
a thermal bremsstrahlung model with $kT=10$ keV gives a similar (0.1-10 keV)
luminosity of $L_{\rm x} = 1 \times 10^{41}$ erg\,s$^{-1}$. 
The $Chandra$ X-ray source is located at RA=09:52:09.56, DEC=21:43:13.3 (J2000)
and agrees well with the SDSS coordinates,
and with the $GROND$ position. 
The astrometric accuracy of $Chandra$ of 1$^{\prime\prime}$ 
corresponds to a projected scale of 1.5 kpc within the host galaxy.

\subsection{$Spitzer$ IR observations} 

In order to search for emission lines excited by the flare, we 
initiated a DDT observation of SDSSJ0952+2143 with the $Spitzer~Space~Telescope$, which
was carried out on 2008 June 5. 
The data also allow us to measure the 
IR SED, get clues on the type of host galaxy,
and measure the post-flare IR luminosity, in
comparison with previous measurements which indicated
possible NIR variability (paper I).
For few, if any, of such extreme transients as SDSSJ0952+2143, MIR 
spectroscopy was ever done.

The observation was carried out with the Short-High module (SH;
wavelength range 9.9-19.6$\mu$m) of the InfraRed Spectrograph (IRS;
Houck et al. 2004). We used the standard staring mode, in which the
target is placed at two nod positions within the slit. 10 spectra
with 120\,s  
duration each were taken at each nod position, amounting
to a total on-source exposure time of 2438\,s. An off-source position
was observed in a similar way to provide a measure of the background.

The raw data were processed by the $Spitzer$ pipeline version
18.0.2 and the residual sky was subtracted from the resulting
two-dimensional spectrum using the off-source observation.
Bad pixels were removed and the areas then Poisson-smoothed. 
Subsequently, 1D spectra for each nod position were extracted from the
Basic Calibrated Data with SPICE
2.1.2\footnote{http://ssc.spitzer.caltech.edu/postbcd/spice.html}. 
The spectra from the two nods were co-added 
and the
final spectrum is shown in Fig.~\ref{fig:irs}.

SDSSJ0952+2143 was detected with a flux density of 12$\pm$1\,mJy 
at 12.5\,$\mu$m. 
The global SED indicates excess
emission around 11\,$\mu$m, and an increase at longer
wavelengths, and closely resembles the mean PG quasar $Spitzer$ IRS SED
of Netzer et al. (2007; their Fig. 3).
We have approximated the MIR SED by two black bodies (with $T_{1}$=460K;
the lower temperature of the second component is not well constrained). 
While this approximation is certainly too simple,   
we use this
parametrization for an estimate of the luminosity.  Integration
over the narrow IRS SH band gives $L_{\rm 10-20\mu m} = 3.5 \times 10^{43}$ erg\,s$^{-1}$,
which is a safe lower limit on the total IR luminosity.
Integrating over the higher-temperature black body component 
gives $L_{\rm MIR} = 9 \times 10^{43}$ erg\,s$^{-1}$. 

We have searched for emission lines
in the spectrum. These lines could either be excited by the flare,
or else represent permanent emission from the host galaxy.   
As a guideline, we have paid attention to emission features
which are commonly detected in the spectra of starburst galaxies, AGN
and supernovae.  Strong MIR lines frequently seen in 
AGN include [SIV]10.51$\mu$m, [NeII]12.81$\mu$m, [NeV]14.32$\mu$m and [NeIII]15.56$\mu$m,
while occasionally [ArV]13.10$\mu$m, [MgV]13.52$\mu$m and [FeII]17.94$\mu$m
have been detected (e.g., Sturm et al. 2002, Netzer et al. 2007, Tommasin et al. 2008,
Dale et al. 2009). The Polycyclic Aromatic Hydrocarbon (PAH; Puget \& Leger 1989)
feature at 11.25$\mu$m is frequently present, too, together with some molecular
Hydrogen lines. In SNe, transitions from Co and Ni have been detected (e.g., Roche et
al. 1989, Kotak et al. 2006).   

The relatively faint spectrum complicates the line searches. 
The enhanced fluctuations in the red part of the spectrum we trace back to
noise.   
There is some evidence for the PAH feature at 11.25$\mu$m,
and emission lines of 
[NeII]12.81$\mu$m and 
[NeIII]15.56$\mu$m.
In order to asses the reality of the Neon lines, we have extracted
spectra from each nod position separately. There is evidence for these
lines in both spectra.  
Other apparent features that appear in the spectrum coincide with regions of 
enhanced background noise. 
No strong emission from [NeV]14.32$\mu$m is detected. 
A more rigorous 
search for faint features is deferred to a future publication. 
Here, we conservatively use the measured flux in [NeIII]
as upper limit for any true line emission, 
which gives $L_{\rm [NeIII]} < 2\times 10^{40}$ erg\,s$^{-1}$.    

\section{Emission-line and continuum diagnostics}

We now derive constraints on the physical
conditions in the reprocessing material, which are later used
in discussing outburst scenarios.

\subsection{Forbidden lines}

Generally speaking, the wide range of ionization states we see,
and the very different critical densities of transitions
among the narrow forbidden lines
(e.g., [SII], [OI] and up to [FeXIV]), imply a wide range of densities
in gas which is kinematically relatively quiescent with
FWHMs of $\sim200-800$ km\,s$^{-1}$.

\subsection{Multi-peaked Balmer-line profile} 

The Balmer-line profile is complex, and variable in time. It consists of
several kinematical components: (1) A broad component with a redshifted peak.
The profile slightly narrowed and the peak shift slightly changed
between 2005 and 2008.  
(2) A narrow component with a width similar to other forbidden lines,
consistent with being constant between 2005 and 2008.
(3) Two peculiar narrow, unresolved, horns, which are present in
H$\alpha$ and H$\beta$ and which faded significantly between 2005 and 2008. 
We comment on each component in turn.

{\it{Broad component, Balmer decrement.}} As already remarked in paper I,
the Balmer decrement is very large, indicating either optical depth effects or a dominant 
contribution from collisional excitation 
(heavy extinction is unlikely, since correction for it would boost the already high Balmer-line
luminosity even further). 
The intensity ratio H$\alpha_{\rm b}$/H$\beta_{\rm b}$ changes from
$\sim12.2$   
to $>5.7$ between 2005 and 2008;
H$\beta_{\rm b}$ is no longer safely detected in the 2008 $NTT$ spectrum. 
The H$\alpha$ luminosity during the 2005 SDSS observation is very large:
$L_{\rm H\alpha_{\rm b}} = 3 \times 10^{41}$ erg\,s$^{-1}$. It decreased by a factor 3
between 2005 and 2008.  
Assuming an average luminosity of 2 $\times 10^{41}$ erg\,s$^{-1}$ lasting for at least 3 years translates into
an energy of at least $2 \times 10^{49}$ erg in H$\alpha$ only.
While the broad H$\alpha$ component may well be completely dominated
by collisional excitation, the narrow components of H$\alpha$ 
and [OIII] are 
still almost as luminous ($L_{\rm H\alpha_{\rm n}} = 2.5 \times 10^{40}$ erg\,s$^{-1}$) 
in 2008 as the detected X-ray 
emission. This either implies 
that the continuum source is partly obscured while the Balmer-line emitting
region is not, or there is a strong EUV contribution to the SED, or we see 
time delay effects of a once brighter continuum.
The broad H$\alpha$ component is redshifted by $\sim$600 km\,s$^{-1}$ in 2005 (paper I), and still by 
300 km\,s$^{-1}$ in 2008. This redshift likely reflects a true kinematic shift. If it
was due to extinction or optical depth effects in a shell geometry,
the receding, more distant, red part of the line would be more affected
(which would cause an apparent blueshift of the centroid).   

{\it{Narrow component.}} This component has similar width than other
narrow lines. The Balmer decrement, H$\alpha_{\rm n}$/H$\beta_{\rm n}$ = 3.2 (2005)
and 2.6 (2008) is close to the Case B recombination value,
indicating little global extinction along the sightline within
the host galaxy of SDSSJ0952+2143.
The luminosity of H$\beta$, $L_{\rm H\beta_{\rm n}}=10^{40}$ erg\,s$^{-1}$,
can be used to estimate the emitting volume
in dependence of gas density and temperature. Under photoionization Case B 
conditions (Osterbrock 1989),
and assuming $T=2\,10^4$ K, a density of $\log n = 2$ as in the [SII] emitting
region (see Sect. 3.4 below)
would imply an emitting volume with a size of $r$=50 pc, 
while $\log n = 4$ implies  $r$=2 pc.
The observed H$\beta$ luminosity further implies a minimum rate $Q$ of H-ionizing photons
of $Q=3.8\,10^{12} L_{\rm H\beta_{\rm n}} = 4\,10^{52}$ s$^{-1}$ assuming full covering
of the continuum source.   

{\it{Narrow horns.}} Apart from the broad base and narrow core, the Balmer lines show
a remarkable extra structure in form of two very narrow, unresolved, horns. 
These horns do not have any counterparts in any other emission lines,
and our new $NTT$ spectrum shows that they do vary between the 2005 SDSS
observation and the 2008 $NTT$ observation, and have become significantly fainter.
The absence of the horns in other emission lines may be a consequence of 
high density.  However, some coronal lines have very high critical density and are still
not detected.  
Alternatively, the lines may form in collisionless shocks. 
These shocks produce strong Hydrogen lines with a two-component profile consisting
of a narrow component contributed by cold Hydrogen atoms, and a broad component
from Hydrogen atoms that have undergone charge transfer reactions with hot protons
while other optical forbidden lines are very faint (e.g., Raymond 1995, Heng \& McCray 2007). 
The flux ratio of broad over narrow component depends on the shock velocity. 
In order to see whether there is extra components in the broad Balmer lines of SDSSJ0952+2143 matching
the narrow horns, higher S/N observations are needed. 
The line shape (double-peaked profile) could arise 
in the case of two colliding streams of gas, in this case 
moving
at a few hundred km\,s$^{-1}$; or alternatively in a two-sided jet or ring morphology
or a bipolar outflow, and will be further discussed below.

\subsection{Coronal lines}

Several flux ratios of emission lines of iron
are of diagnostic value, since they are temperature and/or density sensitive
(e.g., Nussbaumer \& Storey 1982, Keenan \& Norrington 1987). 
The ratio [FeVII]3759/[FeVII]6087 
was used in paper I for a first estimate
of the temperature 
of the emitting gas.
The ratio reported in paper I
was based on total line fluxes. Here we have decomposed both
lines into a broad and a narrow component (Sect. 2.1.4) and 
remeasured the line ratio, which gives [FeVII]3759/[FeVII]6087=0.9
for the broad component and [FeVII]3759/[FeVII]6087=0.7 for the narrow
component in 2005. These ratios imply gas temperatures in the range $T=20000-50000$ K
for the gas which emits the broad component, and $T=15000-30000$ K 
for the gas which emits the narrow component (Keenan \& Norrington 1987).  
[FeVII]3759
could not be decomposed into two components in 2008.
The inferred temperatures are consistent with a photoionization origin
of the emission lines.  Further, the widths of the narrow iron-line cores, of order
$\sim$200 km\,s$^{-1}$ or less, argue against direct collisional ionization by
shocks, since a minimum shock velocity of 300 km\,s$^{-1}$ is required to
collisionally ionize Fe$^{10+}$ (Viegas-Aldrovandi \& Contini 1989).
This
statement does of course not exclude the presence fast 
radiative shocks (Sutherland \& Dopita 1995, Dopita \& Sutherland 1996)
which produce local ionizing radiation which can then photoionize low-velocity pre-shock gas.
In that case, the degree of ionization would be higher in the ambient 
pre-shock gas than in the cooling post-shock gas. In situ production of the ionizing
radiation also ensures efficient reprocessing into emission lines.           

The ratio [FeVII]5158/[FeVII]6087 = 0.2
in 2005 implied an electron density $\log n_{\rm e} = 6-7$,
while its decrease to [FeVII]5158/[FeVII]6087 $<$ 0.06 in 2008 implies
a higher density $\log n_{\rm e} > 7$ (Keenan \& Norrington 1987){\footnote{note
remaining uncertainties in the collision strengths of [FeVII]; e.g. Sect. 3.3 of Ferguson et al. 1997.}}. 
Further, it is reasonable to assume that the bulk of the iron emission lines
should be emitted from gas below the critical density, which is 3\,10$^{7}$ cm$^{-3}$
for [FeVII]6087, and $\sim$ 10$^{9}$ cm$^{-3}$ for [FeX] and [FeXI].

The 2008 line luminosity of [FeVII]6087, $1.6 \times 10^{40}$ erg\,s$^{-1}$, is
almost as high as the simultaneously measured soft X-ray luminosity,  
implying that we either only see a fraction of
the produced soft X-ray luminosity (the rest being completely absorbed),
or that a strong EUV excess is present, 
or else that we still see the echo from a once brighter
flare. Since the recombination timescale in the high-density coronal-line-emitting
gas is very short (at $T=2\,10^4$ K and $\log n=7$,
the Hydrogen recombination time scale is on the order of 10\,d), 
light travel time effects would then have to play a role.  

\subsection{[SII] ratio}

The emission-line intensity ratio [SII]6716/[SII]6731 is 
sensitive to density (Osterbrock 1989, Dopita \& Sutherland 2003).
The measured ratio, 1.3 in 2005 and 1.1 in 2008, indicates
densities of $n \simeq 1\,10^{2}$ cm$^{-3}$ and $n \simeq 3\,10^{2}$ cm$^{-3}$, 
respectively (at T=10$^{4}$ K).   
[SII] therefore signals gas of much lower density than
indicated by the coronal lines.

\subsection{The strength of [OIII]4363/5007}

The great strength of [OIII]4363/5007 $\simeq 0.2-0.3$ in 2005 and 2008  
significantly exceeds the value typically
observed in Seyfert galaxies and is above
photoionization predictions for a large parameter space (e.g., Fig. 9 
of Komossa \& Schulz 1997, Groves et al. 2004).
The observed value implies that we
are beyond the low-density regime so that the ratio no longer only depends on
temperature, but also becomes 
density sensitive (Osterbrock 1989, Dopita \& Sutherland 2003). 
For a temperature of  $T = (2-5)\,10^{4}$ K the ratio
implies a density 
on the order of $\log n = 7$ (Dopita \& Sutherland 2003).

\subsection{NIR emission}

Given that the 2MASS measurements most likely represent the host galaxy  
emission {\em prior} to the outburst
we have used the NIR colours to determine
the nature of the host galaxy. The values of J-H=0.9 mag and H-K$_{\rm s}$=0.1 mag locate 
SDSSJ0952+2143 
in the area typically populated by non-active galaxies in two-colour diagrams 
(e.g., Hyland \& Allen 1982, Willner et al. 1984, Glass \& Moorwood 1985, Alonso-Herrero et al. 1996; our Fig. 6).

\subsection{MIR SED}

The global $Spitzer$ SED of SDSSJ0952+2143 indicates excess
emission around 11\,$\mu$m, and an increase at longer
wavelengths, and closely resembles the mean PG quasar $Spitzer$ IRS SED
of Netzer et al. (2007; their Fig. 3).
The bumps in those quasar spectra are traced back to silicate emission
(see also Schweitzer et al. 2006, and references therein).
The extrapolation of the observed $Spitzer$ MIR SED falls 
well above the NIR emission (dominated by the extended host) by an order of magnitude (Fig. 8),
implying that the IR emission of SDSSJ0952+2143 
is completely dominated by relatively cold dust -- unlike
quasar spectra which commonly show a significant emission component around a few $\mu$m
from relatively hot dust, heated by the radiation from the AGN accretion disk.  
The observed MIR emission is extraordinary luminous, 
with $L_{\rm 10-20\mu m} = 3.5 \times 10^{43}$ erg\,s$^{-1}$
in the narrow IRS SH band. 
The relatively low mass of the SMBH of the host galaxy of SDSSJ0952+2143,
$M_{\rm BH} \simeq 7\,10^{6}$ M$_{\odot}$ (paper I; see our Tab. 3), 
implies that the observed
10-20$\mu$m luminosity is already 1/25 of the Eddington luminosity.   

With only one MIR spectrum at hand and no imaging information, 
there is currently two possibilities
to explain the MIR emission. Either we see {\em temporary} emission from
dust heated by the flare, or else this is {\em permanent} emission from a
starburst region. The observed MIR luminosity is comparable to that of
nearby IR selected Seyfert galaxies (e.g., Tommasin et al. 2008), so quite luminous.
At the same time, we do not see evidence for a young stellar population
in the optical spectrum. So, if the MIR emission was permanent, the starburst region would
then have to be completely obscured in the optical band. 

Within the limits of the black body approximation, we 
can estimate  
the size of the emission region.
For the given black body temperature and MIR luminosity, we
obtain a radius of the emission region of 0.5 pc. 

We do not have tight constraints on the geometry of the dust distribution. 
However, the lack
of reddening of the narrow Balmer lines implies that the dust is
not in a spherical shell around this line-emitting region;
and the strength of the coronal lines implies that this region
is free of silicates, otherwise Fe would be heavily depleted onto dust grains.

\section{Discussion} 

\subsection{Outburst mechanisms}

The strong emission-line and the low-energy continuum variability of SDSSJ0952+2143
are the consequence of  
a high-energy outburst which was itself not observed. The emission-line
and continuum variability encompassing several wavebands
and forbidden and allowed emission lines, are unique
among variability associated with the cores of galaxies.  

Several different outburst mechanisms can potentially produce
the very unusual emission-line signatures of  
SDSSJ0952+2143. We briefly
introduce them in this section, and then come back to each of
them in more detail in the following subsections. 
(1) An accretion disk around a SMBH will produce ionizing radiation
that is reprocessed by surrounding gas clouds into characteristic broad
and narrow emission lines which are commonly observed in the spectra of
AGN (Osterbrock 1989). Changes in the accretion
rate for instance in form of a disk instability could lead to highly variable continuum emission
and excite the unusual emission lines (Sect. 4.2). 
(2) A flare of electromagnetic radiation can be produced from the
temporary accretion disk formed by the debris of a star 
tidally disrupted by a supermassive black 
hole in the core of SDSSJ0952+2143. As the ionizing radiation
travels through the nucleus, emission lines originate in the
ISM and also in part of the stellar debris itself (Sect. 4.4). 
(3) A fraction of supernovae emit significant X-ray radiation, 
which likely originates in shocks driven into the circumstellar medium (CSM) by the SN ejecta. 
This radiation ionizes ambient gas, including the SN ejecta, the
(clumpy) progenitor wind and the ISM. These components then emit broad and sometimes
also narrow emission lines (Sect. 4.3). 
A small fraction of SN spectra look surprisingly similar to those of AGN
(e.g., Fig. 6 of Filippenko 1989, Fig. 1 of Terlevich \& Melnick 1988), 
even though essentially all AGN
and SN spectra look very different from SDSSJ0952+2143.

Finally, there is a region of parameter space where the mechanisms
(1)-(3) discussed above  overlap;
including SNe exploding in gas-rich AGN cores, stellar tidal detonations,
and flung-out stellar tidal debris interacting with the ISM (Sect. 4.5).

\subsection{AGN outburst}

Some fraction of all AGN show coronal line emission in their optical spectra
(e.g., Seyfert 1943, Penston et al. 1984, Oliva et al. 1994, Binette et al. 1997, Nagao et al. 2000,
Mullaney \& Ward 2008). 
The large majority of these coronal lines are constant or, very rarely, show
slight variability (e.g., Netzer 1974, Veilleux 1988).  
Only one, the active galaxy IC3599 (Brandt et al. 1995, Grupe et al. 1995, Komossa \& Bade 1999),
has shown strong transience in its coronal lines. 
In response to a luminous X-ray flare, several bright emission lines appeared and then faded on the
timescale of years. The total amplitude of variability is large 
and the X-ray spectrum is very soft, both in 
high-state and low-state (Brandt et al. 1995, Grupe et al. 1995, Vaughan et al. 2003).
A change in accretion rate (perhaps due to a disk instability or due to stellar tidal disruption)
is a likely explanation of these observations (Brandt et al. 1995, Grupe et al. 1995).    

Clearly, with its strong coronal lines
and dramatic line variability, SDSSJ0952+2143 is different from
AGN in general, and is more extreme than IC3599 in its broad-band 
emission-line and continuum response.

If a high-amplitude outburst occurred in an AGN, different emission-line regions -- 
the broad line region (BLR), coronal line region (CLR) and 
narrow line region (NLR)  -- would be illuminated at different times, and will respond with
different recombination timescales. Consequently, ``outburst spectra'' 
could look significantly different from equilibrium spectra. 
The range of densities inferred from the emission-line
spectrum of SDSSJ0952+2143 is consistent with 
indications for local density inhomogeneities in
the classical NLR of AGN (e.g., Komossa \& Schulz 1997, Brinkmann et al. 2000). 

As noted in paper I, the ratio [OIII]/H$\beta_{\rm n} \simeq 3$ formally
places SDSSJ0952+2143 at the border between AGN and LINERs
in diagnostic diagrams; but given the variability of most
or all emission lines, this cannot at all be used as a classification
of the galaxy.{\footnote{If we momentarily assume that the [OIII] emission is permanent and
use the known correlation between [OIII] luminosity and X-ray luminosity among AGN 
of Heckman et al. (2005, see also Netzer et al. 2006), we
predict a (2-10) keV X-ray luminosity  of $L_{\rm 2-10} \sim 10^{42}$ erg\,s$^{-1}$;
higher than observed by more than one order of magnitude. }}   
In terms of emission-line ratios, the classification
of SDSSJ0952+2143 as active or inactive galaxy in quiescence therefore 
remains unknown at present. 

However, the pre-flare 2MASS colours indicate a non-active
galaxy, the low-state X-ray flux measured with $Chandra$ does not hint at 
the presence of a permanent AGN (except possibly a low-luminosity AGN), 
and the optical spectrum barely
has a detectable non-stellar component.
We therefore consider the presence of a classical AGN at the core of SDSSJ0952+2143,
which underwent an unusual outburst, very unlikely.

\subsection{Supernova explosion} 

\subsubsection{Supernovae of type  IIn} 

One subclass of supernovae, those of type IIn (Schlegel 1990)
show narrow emission lines preferentially from Hydrogen and Helium  
in their optical spectra in addition to broad lines
(see Filippenko 1997 for a review; his Fig. 14){\footnote{SNe of this type are,
e.g., SN 1987F (Wegner \& Swanson 1996),  
SN 1988Z (Stathakis \& Sadler 1991, Turatto et al. 1993, Aretxaga et al. 1999), 
SN 1995G (Pastorello et al. 2002), SN 1995N (Fransson et al. 2002), SN 1994W (Chugai et al. 2004),
SN 1997eg (Hoffman et al. 2008), SN 1998S (Leonard et al. 2000),  
SN 2005gl (Gal-Yam et al. 2007),
SN 2006tf (Smith et al. 2008a),   
and 
SN 1997cy (Germany et al. 2000, Turatto et al. 2000).
We note in passing that SNe associated with GRBs
are typically of type Ic which lack Hydrogen features in their optical spectra.}}.   
These lines are interpreted as a result of strong interaction
of the SN ejecta with dense CSM (e.g., Chugai 1990, Chugai \& Danziger 1994).
Only a few SNe are known which exhibit coronal lines 
(e.g., SN 1988Z: Turatto et al. 1993,
SN 1993J: Garnavich \& Ann 1994, SN 1995N: Fransson et al. 2002, SN 1987A: Gr\"oningsson et al. 2008,
SN 1997eg: Hoffman et al. 2008, SN 2005ip: Smith et al. 2009)
and detection of ionization states as high as [FeXIV] is very rare. 
According to Smith et al. (2009), SN 2005ip is unprecedented in the way the coronal
lines dominate the optical spectrum of that SN. 

Several SNe IIn are also relatively X-ray luminous, 
exceeding 10$^{40}$ erg s$^{-1}$
(e.g., SN 1988Z: Fabian \& Terlevich 1996, Aretxaga et al. 1999, 
Schlegel \& Petre 2006; SN 1986J: Bregman \& Pildis 1992, Temple et al. 2005; 
SN 2006jd: Immler et al. 2007; see   
Schlegel 1995, 2006, and Tab. 1 of Immler 
2007{\footnote{see 
http://lheawww.gsfc.nasa.gov/users/immler/supernovae\_ \\
list.html maintained by S. Immler for updates}}
for overviews).
Strong ionizing radiation may originate from the shock break-out,
Compton-upscattering, and in shocks when SN ejecta collide
with the precursor wind and/or the ISM.
A number of type IIn SNe show excess emission in the NIR, plateaus in their NIR
lightcurves, and characteristic variability of their Balmer-line profiles. 
These properties have been traced back to the presence of dust; either pre-existing
dust which is heated collisionally or radiatively after the SN explosion, 
or newly formed dust which coagulated after the SN explosion in cool, dense regions
(e.g., Merrill 1980, Lucy et al. 1989, Roche et al. 1989, Moseley et al. 1989, Wooden et al. 1993,
Gerardy et al. 2002, Pozzo et al. 2004, Sugerman et al. 2006, Meikle et al. 2006,
Smith et al. 2008b, Fox et al. 2008 and references therein). 

There are several key sites of emission line formation.
The radiation produced in the shocks will ionize the surrounding ISM, the stellar wind
of the SN precursor and the high-density SN ejecta themselves.
The narrow
lines likely arise in the CSM (sometimes in the ISM); while broad lines likely
come from the SN ejecta (e.g., Chugai \& Danziger 1994, Filippenko 1997).
Generally, SNe of type IIn show substantial heterogeneity with respect to
their luminosities and emission-line properties.
They are relatively rare, comprising  only
a few percent of all core-collapse SNe (Gal-Yam et al. 2007, Smartt et al. 2009). 

\subsubsection{Super-luminous iron lines, and comparison with SN 2005ip and SN 1988Z} 

Given similarities of the optical spectra of SDSSJ0952+2143 first
presented in paper I with the recently published remarkable spectra
of the type IIn supernova SN 2005ip (Smith et al. 2009; S09 hereafter), 
we consider an origin of SDSSJ0952+2143 
in terms of a similarly unusual supernova. 
S09 reported very strong coronal
lines including a large number of transitions, 
and ionization states up to [FeXIV], with a peak luminosity (at day 93) in
[FeXIV] of $1.5 \times 10^{38}$ erg\,s$^{-1}$.
SN 2005ip and several other type IIn SNe also
show high ratios of [OIII]4363/[OIII]5007,
broad Balmer lines with widths of at least (one -- several) thousand km\,s$^{-1}$ 
and very high Balmer decrements (up to 25 in the case of SN 2005ip).
The peak luminosity of SN 2005ip was only modest with an (unfiltered)
absolute magnitude of --17.4 (S09).  

SDSSJ0952+2143 is phenomenologically similar in several 
of its emission-line properties, 
especially line ratios, but much
brighter in the optical, and brighter in X-rays, and, in particular,
much more luminous in the coronal lines. 
According to S09, SN 2005ip is the SN with the strongest
iron coronal lines, in the way they dominate the spectrum of that SN.
The luminosity in the high-ionization coronal
lines of SDSSJ0952+2143 is a factor of $\sim$ 100 higher (!) than the
highest value measured in SN 2005ip (S09)
and in SN 1988Z (Turatto et al. 1993) and it is a factor 
10$^{3}$ higher than in SN 1995N (Fransson et al. 2002),
and a factor 10$^{4}$ higher than in SN 1993J (Garnavich \& Ann 1994).{\footnote{Note that only
few spectra of SDSSJ0952+2143 exist;
its peak line luminosity could have therefore been even higher. }} 

In the context of an SN interpretation, SDSSJ0952+2143
therefore is the SN with the most luminous iron coronal lines. 
SDSSJ0952+2143 is also among the very luminous SNe II
in terms of continuum emission
(c.f. Fig. 3 of Smith et al. 2008a), with an R magnitude
of --20.8 mag in 2004 (including emission from the host galaxy, but likely
dominated by the transient),
and it almost
rivals the ``hypernova'' SN1997cy (Turatto et al. 2000, Pastorello et al. 2002) 
in its large H$\alpha$ luminosity.
In the context of the SN scenario, 
the density of the forbidden line-emitting gas of SDSSJ0952+2143, 
$\log n \simeq 2...7$, is consistent
with a clumpy progenitor wind and/or ISM (but could also imply SN explosion
into a star-forming region), the lowest-velocity gas with
the pre-shock wind speed of an LBV progenitor (e.g., Humphreys \& Davidson 1994, Smith et al. 2008a). 

On the other hand, there are also significant spectral 
differences between SDSSJ0952+2143 and
SN 2005ip (and other similar SNe), for instance:
(1) No HeI transitions are detected in our spectra;
neither broad nor narrow, whereas they are frequently seen in SNe. 
(2) The narrow horns in the Balmer
lines of SDSSJ0952+2143 are unusual (but see below). 
(3) The profile asymmetry of H$\alpha$ of
SDSSJ0952+2143 is opposite to that  
sometimes seen in SNe (which show apparent blueshifts; often traced back to 
stronger extinction in the red part of the line from the far side of the SN ejecta; e.g., 
Lucy et al. 1989, S09). 

Regarding the narrow horns in the Balmer lines, it is interesting to note
a potential similarity with SN 1998S, which
showed a double-horned profile (Garnavich et al. 1998), traced back 
by Gerardy et al. (2000) to the presence of interacting
dense clumpy circumstellar gas with a ring-like morphology. 
However, they measured the three peaks at velocities of -4900, -400 and +3300 km\,s$^{-1}$;
at much wider velocity separation than the peaks of SDSSJ0952+2143. The lines are also
much broader in SN 1998S. 

\subsubsection{Luminous X-ray emission} 

$Chandra$ indicates a relatively hard X-ray spectrum of SDSSJ0952+2143, frequently seen in SNe
(thermal spectra with temperatures of several--10 keV).
The observed X-ray luminosity of SDSSJ0952+2143 is higher than that of most other known SNe.
Only $\sim$40 SNe are X-ray detected at all
(according to Tab. 1 of Immler 2007, and adding recent detections)
and only a handful of them are more luminous than 10$^{40}$ erg\,s$^{-1}$.
One of the most luminous (long-duration) X-ray emitting SNe is SN1988Z (Fabian \& Terlevich 1996),
at $L_{\rm x} \approx 10^{41}$ erg\,s$^{-1}$. 
SN 2005ip was detected at $1.6 \times 10^{40}$ erg\,s$^{-1}$
(0.2-10 keV, at day 461; Immler \& Pooley 2007). 
If of SN origin, SDSSJ0952+2143 is one of the most distant X-ray detected, and one of
the most X-ray luminous
SNe known. 

Perhaps the tightest constraints on any SN-based outburst scenario
are set by the huge luminosities in the narrow emission lines.
The luminosity in the highest ionization coronal lines implies
very luminous intrinsic soft X-ray emission under photoionization conditions.
So far, line luminosities $>10^{40}$ erg\,s$^{-1}$ as in SDSSJ0952+2143
have only been observed in AGN (e.g., Nagao et al. 2000, Mullaney \& Ward 2008). 
While the ionization potentials of [FeVII]-[FeXIV] (0.1-0.4 keV) fall in the X-ray
regime, it is not straight-forward to convert the observed coronal 
line luminosities into X-ray luminosity, since two processes contribute
to the line excitation, collisional excitation and resonance fluorescence by UV photons
(e.g., Osterbrock 1969), and since atomic parameters are still
relatively uncertain (e.g., Pelan \& Berrington 1995, Mohan et al. 1994, Oliva 1997, Berrington 2001). 
In AGN, the observed coronal line luminosity is approximately
0.1\% (-1\%) of the observed X-ray luminosity. If the line formation
efficiency was similar under SN conditions, in the case of SDSSJ0952+2143
this would imply an intrinsic soft X-ray luminosity on the order of 
at least $\sim$10$^{42-43}$ erg\,s$^{-1}$. 
No such X-ray luminous SN has ever been observed (or identified as such)
[the luminous X-ray emission of SN 2008D only lasted for seconds-minutes;
Soderberg et al. 2008]. 
An SN explosion into a dense medium (Wheeler et al. 1980, Shull 1980) 
was invoked to explain the luminous X-ray emission of SN 1988Z
(Fabian \& Terlevich 1996), but it is very difficult to reach
luminosities of 10$^{42-43}$ erg\,s$^{-1}$ by this mechanism (Komossa \& Bade 1999).

\subsubsection{Luminous IR emission}

Few SNe have MIR measurements and spectroscopy
in particular (e.g., Kotak et a. 2006, Gerardy et al. 2007).
SN 1987A is the best-studied exception (e.g., Roche et al. 1989, Moseley et al. 1989,
Wooden et al. 1993). Its MIR SED is relatively flat, indicating that
it is dominated by an almost featureless dust component (Wooden et al. 1993),
perhaps graphite (Moseley et al. 1989).
The characteristic structure
of the $Spitzer$ spectrum of SDSSJ0952+2143
indicates the presence of silicates.
The MIR emission is extraordinarily luminous, with at
least $L_{\rm 10-20\mu m} = 3.5 \times 10^{43}$ erg\,s$^{-1}$.
This compares to other IR luminous SNe which only reached
up to several 10$^{40-41}$ erg\,s$^{-1}$ in the NIR (e.g., Gerardy et al. 2002, Pozzo et al. 2004),
while the most luminous SNe ever observed reached peak luminosities
of few $\times 10^{44}$ erg/s in the optical band 
(e.g., Quimby et al. 2007).

The black body radius inferred from the MIR
emission of SDSSJ0952+2143 (Sect. 3.7) corresponds to 0.5 pc.
This is much larger than the distance, the SN ejecta could have travelled within several
years and implies that we see an infrared light echo of pre-existing dust.
The huge MIR luminosity raises the question whether we instead see permanent
emission from a dusty star-forming region.
We cannot entirely rule out this possibility, but it is unlikely because
a luminosity of similar magnitude was inferred from the presence of
the [FeVII] coronal lines in the previous Section, and this (UV--X-ray)
radiation is a plausible heating source of the dust.
We therefore expect that the dust emission maximum continues
to shift further towards the FIR as the dust cools and the incident radiation fades away.
A new IR observation with $Spitzer$ or the $Herschel~Space~Observatory$
would allow us to check whether this trend indeed exists.

\subsubsection{Location and environment}

A dense environment of SNe of type IIn is usually invoked to explain the   
luminous continuum radiation of several of these SNe, and is also
inferred from emission-line characteristics. The densities we derive 
are consistent with such an interpretation. A dense environment of
SDSSJ0952+2143 is also consistent with the strong MIR light echo
from dust, and with the location of the transient within the
galaxy core where dense gas is plausibly most abundant. 
At the same time, the blueness of the optical powerlaw continuum 
and the $GALEX$ colours argue against heavy extinction 
along the line of sight. This would then locate the event at the edge of
a molecular cloud complex, where most dust directly along our
line of sight was destroyed by the bright flare.
Finally, we note that the evidence for spiral structure in the
$GROND$ images is not inconsistent with
a SNII interpretation; type IIs have not
yet been observed in elliptical galaxies (Capellaro et al. 1999).

\subsection{Transient accretion event in a non-active galaxy} 

Given the extreme luminosity requirements in the context of an SN scenario,
we now return to accretion power as energy source. 
Since there is no evidence for classical permanent AGN activity in
SDSSJ0952+2143 in the form of a permanent optical and X-ray bright accretion disk,
we focus now on the possibility of a temporary accretion event in a non-active
galaxy, for instance by accretion of a molecular cloud or
the tidal disruption of a star. These events could similarly
happen in a low-luminosity AGN the presence of which
we cannot yet exclude. In the latter, an accretion disk
instability could also temporarily enhance the accretion luminosity.  
That said, we mostly concentrate on the tidal disruption scenario in what follows. 

Interestingly, all of the observed emission-line and continuum
properties can also potentially be understood in the context of a stellar tidal 
disruption event; where a star is disrupted in a gas-rich 
core environment.

\subsubsection{Tidal disruption of a star by a SMBH}

A star approaching a SMBH will be tidally disrupted once the tidal
forces of the SMBH exceed the star's self gravity. A fraction of the 
stellar debris will subsequently be accreted, producing a luminous flare
of radiation (e.g., Hills 1975, Young et al. 1977, Gurzadian \& Ozernoi 1981, Luminet 1985,
Rees 1988, Evans \& Kochanek 1989, Laguna et al. 1993, Li et al. 2002,
Gomboc \& Cadez 2005, Ivanov \& Chernyakova 2006;    
see Komossa 2002 for a much longer list of references on theoretical aspects 
of tidal disruption) peaking in the X-rays or UV and lasting on the order of months
to years. 

X-ray variability, up to factors of $\sim$6000,
has been almost exclusively detected in non-active
galaxies, and these events represent the best observational evidence to date for
the process of tidal disruption of stars by SMBHs
(e.g., Komossa \& Bade 1999, Halpern et al. 2004,
Komossa et al. 2004).
The tidal disruption candidates reached huge X-ray luminosities
up to at least $L_{\rm sx} \approx 10^{44}$ erg\,s$^{-1}$
and show a lightcurve which jointly declines as
$t^{-5/3}$, as predicted in some variants of tidal disruption models (e.g., Rees 1990).
A few events in the UV shared similar properties with these X-ray events
[similar decline law based on well-covered light curves, possibly soft X-ray spectrum,
and (much lower) UV amplitude of variability of a factor of a few] 
and were thus interpreted along the same lines (Gezari et al. 2008; see also
Renzini 2001, Stern et al. 2004).
None of the tidal disruption flares so far observed from non-active galaxies had
an emission-line ``light echo'' detected, but optical follow-up spectroscopy has typically
only been obtained several years after the flare (Komossa \& Bade 1999, Komossa \& Greiner 1999,
Grupe et al. 1999, Greiner et al. 2000, Gezari et al. 2003, 2008, Esquej et al. 2008, Cappelluti et al. 2009).{\footnote{A few
events of unusual broad line variability in AGN and LINERs have sometimes been linked to accretion
events including tidal disruption; for instance the transient broad double-peaked Balmer lines 
of NGC\,1097 (Storchi-Bergmann et al. 1995, Eracleous et al. 1995, Storchi-Bergmann et al. 2003) and the transient
broad HeII line of NGC\,5548 (Peterson \& Ferland 1986). We also note that faint high-ionization [OIII]
emission detected in the post-flare $HST$ spectrum of NGC\,5905 (Gezari et al. 2003) could have
well been excited by the flare, rather than representing permanent low-level activity.}}  

Apart from the accretion of stellar debris which causes the powerful X-ray flares, 
in the course of a tidal disruption event, ionizing radiation
will also be released in shocks during 
stream-stream collisions (Kato \& Hoshi 1978, Rees 1988, Lee et al. 1996), and 
during collisions of the ejected stellar debris with the
ISM (Kokhlov \& Melia 1996, Ayal et al. 2000).

\subsubsection{Emission-line light echoes} 
 
The radiation from accretion and other 
processes will ionize any surrounding gaseous
matter in the galaxy core:
(1) the interstellar medium of the host galaxy,
including the gas clouds of BLR, NLR, CLR  and torus, if present, 
and (2) the
gaseous debris of the star itself 
which would
be of very high density, show a spread in densities,
have asymmetric spatial distribution,
and may well have an unusual chemical composition.

A tidal disruption flare would therefore enable an extreme
application of ``reverberation mapping'' where all gaseous material responds
to one single flare (Blandford \& McKee 1982) rather than a complicated lightcurve. 
This is of special interest with application to the cores of active galaxies, for instance
to map the response of the molecular torus. It is also of interest with respect to non-active
galaxies, because it would tell us which of those gaseous components typically detected
in active galaxies (like the BLR) are still present in non-active galaxies.
Once applied to a large number of galaxies, this method might also provide a new
route of addressing the nature and velocity field of the gas that fuels
SMBHs.  

Predictions of how emission-line spectra excited by
tidal disruption flares should look like in detail, do not yet exist
and are complicated by the time dependence of the ionization;
early spectra would be dominated by high-ionization emission lines, while 
lines like HeII and some low-ionization lines would
persist especially long (e.g., Binette \& Robinson 1987)
after the flare, when [OIII]5007 has already faded. 
{\footnote{Photoionization modelling of the coronal lines of the flaring galaxy IC3599 has shown
that gas with a density of 10$^9$ cm$^{-3}$ and a column density of 10$^{23}$ cm$^{-2}$ (conditions
similar to the outer BLR or CLR in AGN) can produce the observed line ratios
(Komossa \& Bade 1999). 
In addition, the strong soft excesses seen in tidal disruption candidates (Komossa 2002)
boost the coronal lines in strength. }} 

After the tidal disruption of a star, streams of tidal debris
would form, spreading over a number of orbits. This would result in complex
multi-peaked Balmer emission-line profiles in the early
phase of disruption (days to months; Bogdanovic et al. 2004). 
After that, line profiles would
change less rapidly, and would represent a superposition of different components
of the tidal debris, which would respond to the time-dependent
illumination from the central accretion disk.    
Simulations of the stellar disruption process (e.g., Evans \& Kochanek, Laguna et al. 1993,
Ayal et al. 2000, Gomboc \& Cadez 2005) have shown that approximately 75\% of the
stellar debris is unbound after first pericentre return. The innermost bound debris
circularizes quickly, while the rest is spread over a number of
eccentric orbits. Such a configuration might explain the broad component in H$\alpha$
and its peak shift on the timescale of years, while the high density of the stellar debris
would explain the large Balmer decrement. At the same time, the line luminosity cannot
easily be explained within this scenario (see Sect. 3.2).   
Stream-stream collisions between returning debris and more tighlty bound gas 
might produce double-horned profiles like the one we observe
(see the discussion on collisionless shocks in Sect. 3.2), 
but we would expect much stronger profile variability 
on the timescale of years than is actually observed (while the narrow horns
of SDSSJ0952+2143 do vary in  intensity, 
the velocity separation of the two peaks did not
change within the errors).  Alternatively, the double-horned emission
could represent narrow streams of unbound debris (viewed from
a special direction, since line centroid shifts are small).   
Other geometries which could produce double-horned profiles
involve a jet or bipolar outflow component; the small line width
and peak separation would require a special viewing angle. 

The unbound stellar tidal debris
will be flung out from the system at high speed and
when interacting with the ISM 
could cause effects similar to a supernova remnant (SNR) but much more
powerful (Khoklov and Melia 1996). 
While the available kinetic energy could be much higher than in
a SN, the mechanism only works efficiently if
the escaping debris forms a narrow stream rather than a broad fan
(see the discussion in Khoklov \& Melia 1996). 

The estimated radius of the infrared light echo, 0.5 pc,
is consistent with the expected location of molecular gas 
in the galaxy core. Based on the black hole mass of the host
galaxy, and applying typical scaling relations known in
active galaxies (Elitzur 2007), the inferred size of the emission region
is comparable to the molecular torus.  Any dust mixed with gas
at distances much closer to the nucleus would have been
destroyed by the bright flare.  The fact that we see silicate emission
rather than absorption features further implies that we have a 
relatively unobscured view into the galaxy. 

\subsubsection{Overall energetics}
Tidal flares might be well powered by Eddington-limited accretion shortly
after the disruption. Assuming a $t^{-5/3}$ decline law 
(e.g., Rees 1990, Komossa \& Bade 1999) for the luminosity
evolution after disruption, and extrapolating backwards in time starting
with the observed X-ray luminosity in 2008, the predicted luminosity
is still below the Eddington luminosity in 2004, consistent with
tidal disruption. Given uncertainties in the decline law and
the EUV part of the continuum, this estimate is
uncertain by at least one order of magnitude.    
We note in passing that the observed X-ray spectrum of SDSSJ0952+2143 is much harder
than the previously observed, very soft, tidal flares (e.g., Komossa \& Bade 1999).

\subsection{Links between Supernovae, tidal disruption, and GRBs}

There might well be a region of parameter space where the
mechanisms discussed in previous Sections are related:
It has long been pointed out, and re-confirmed in recent studies, 
that high penetration factors of stars approaching SMBHs could  
ignite nuclear burning in the stellar core (including especially white dwarfs) 
and thus trigger an SN explosion{\footnote{Most of these events would occur 
strictly in galaxy {\em cores} even though tidal disruption/detonation
of stars around recoiling SMBHs (Komossa \& Merritt 2008) and white dwarf
tidal detonation by intermediate-mass BHs (Rosswog et al. 2008)
could happen offnuclear.}} (e.g., Carter \& Luminet 1982, Bicknell \& Gingold 1983, 
Luminet \& Marck 1985,
Luminet \& Pichon 1989,   
Frolov et al. 1994, Dearborn et al. 2005, 
Brassart \& Luminet 2008, Rosswog et al. 2008). In that case, the collisions of
the ejected stellar debris with the ISM could be even more powerful than in a classical
SNR (Khoklov \& Melia 1996). Variants of tidal disruption and detonation
have also been employed to explain certain classes of GRBs (e.g., Carter 1992, 
Lu et al. 2008, Brassart \& Luminet 2008).  Among the GRBs, a few appear to be 
associated with SNe (of type Ic) (e.g., SN 1988bw/GRB980425: Galama et al. 1998, 
SN 2003dh/GRB030329: Hjorth et al. 2003, SN 2006aj/GRB060218: Mazzali et al. 2006;
see Woosley \& Bloom 2006 for a review). 

We have checked the data base{\footnote{http://www.mpe.mpg.de/$\sim$jcg/grbgen.html}}
of well-localized (better than 1 degree) 
GRBs since 1997
in order to see whether any GRB was recorded from the direction of 
SDSSJ0952+2143. This includes GRB detections by HETE II, BeppoSAX,
INTEGRAL, $Swift$, AGILE, Fermi-LAT and the Interplanetary Network.
None was found within 10 degrees. However, we note that the temporal 
coverage of the field was below about 25\%.
SN (in dense media) and GRB-related models have also been
discussed by Komossa \& Bade (1999) for the tidal disruption candidate 
NGC\,5905, but were not favored.  

Another interesting phenomenon is a SN exploding in an ``AGN-like'' gaseous core, e.g., 
in or near a BLR or NLR cloud, or the molecular torus.  
This would complicate emission-line modelling,
and if not identified as such could lead to wrong conclusions
about the properties of the wind of the progenitor star.

At present, the similarities in several spectral properties of
SN 2005ip and SDSSJ0952+2143 suggest a similar mechanism at work;
and the overall appearance of SN 2005ip suggests a classical
SN explosion, albeit with a number of peculiarities.

While it is important to keep in mind potentially large similarities
between SN ``light echoes'' and tidal disruption ``light echoes'',
at present the SN scenario perhaps is the
most conservative interpretation of SDSSJ0952+2143 -- even though 
the coronal line luminosity and the MIR luminosity 
is spectacular and unprecedented.

Other very peculiar iron coronal lines in AGN 
might have an SN origin, too
(while the bulk of AGN coronal lines are spatially extended and have
properties which are best explained by classical models of the coronal line region).

\subsection{Event rate}
In order to estimate the rate of events as unusual as SDSSJ0952+2143,
we have started a systematic search for similar objects
among 7\,10$^{5}$ 
galaxies of the SDSS galaxy data base. A first quick-look study
produced at least two more events which fulfil the following three criteria:
(1) [FeVII]-[FeXI] comparably strong as [OIII]5007, (2) large Fe line luminosities,
and (3) detection of Fe line variability in our follow-up optical
spectroscopy (Zhou et al., in prep.; Komossa et al., in prep.).  All three
events arise from the galaxy cores within the errors. 
There may be more such objects, but taking those 3 safely identified so far
results in a {\em lower limit} on the event rate of 4\,10$^{-6}$/galaxy/(``a few'' years).
The exact value of ``a few'' depends on the longevity of the iron features
which could be on the order of 0.5-5 years.   
This rate compares to theoretical predictions
of the tidal
disruption rate of $\sim$10$^{-5}$/galaxy/year (Merritt 2009), and to observational
constraints on this rate which are on the same order (Donley et al. 2002, Esquej et al. 2008).

\subsection{Future follow-up observations} 
A number of future observations are of interest in further distinguishing 
between different scenarios for SDSSJ0952+2143. 
Several type II SNe turned out to be luminous 
radio emitters (e.g., Weiler et al. 1990, van Dyk et al. 1993),
while in the context of stellar tidal disruption it is still unknown whether
these events will produce significant radio emission. Radio follow-ups are
therefore of interest.  
X-ray monitoring will tell the long-term decline law. While all
tidal disruption candidates showed a continuous fading of their X-ray lightcurves
(Komossa 2002),
supernovae have more complicated X-ray lightcurves which can strongly increase or decrease
with time (Schlegel 2006, Immler 2007). 
A new $GALEX$ or $Swift$ observation should confirm the fading
of the UV emission.
An optical image of sub-arcsecond resolution would
allow accurate localization of the faint continuum source within the galaxy core. 
Optical spectroscopic monitoring of the broad and narrow emission lines  
will provide the best diagnostic of the line-emitting gas.  
Among all spectral features of SDSSJ0952+2143 three stand out as special:
(1) The narrow horns in the Balmer lines; luminous,  with small kinematic separation,
and very small line widths ($< 200$ km\,s$^{-1}$). High S/N observations would much better resolve
the profile, and its temporal evolution. (2) The ultra-luminous coronal lines. [FeVII]6087
is still luminous and persistent in the most recent spectra, and measurement
of its longevity will give tight constraints on the power source of the line emission,
and therefore on the outburst mechanism. (3) The huge MIR luminosity of at
least $3.5 \times 10^{43}$ erg\,s$^{-1}$. One more $Spitzer$ or a $Herschel$ observation would
tell whether this emission is permanent or varies in luminosity and temperature,
and is therefore definitely related to the flare.       

\section{Summary and conclusions} 

After the initial detection of the exceptional optical emission-line spectrum
of SDSSJ0952+2143 and its variability (paper I), 
and suspecting
a tidal disruption event, peculiar AGN or supernova explosion, 
we have carried out multi-wavelength
follow-up observations, employing $Spitzer$, $Chandra$, $NTT$,
the $Xinglong$ telescope, and $GROND$, 
in combination with archival observations with $ROSAT$, $XMM-Newton$, $Swift$, 2MASS, NVSS,
and $GALEX$. 

SDSSJ0952+2143 shows the following continuum
and emission-line properties: 
\begin{itemize} 
\item The 2MASS detection (likely pre-flare) indicates a non-active
galaxy. 
\item The observed X-ray luminosity of $10^{41}$ erg\,s$^{-1}$ in 2008 is below
that of a typical AGN and above a typical SN.
\item  The optical spectrum of SDSSJ0952+2143 is dominated by luminous coronal lines,
and those of the highest ionization have dramatically faded between 2005 and 2008.
\item  The wide range in ionization states and critical densities
of the forbidden emission
lines and several density-sensitive line ratios indicate
a range in electron densities of the line-emitting gas ($\log n = 2..7$).
The temperature-sensitive line ratios imply photoionization
as dominant ionization mechanism.
\item   The luminosity in each of the high-ionization 
lines [FeVII]6087, [FeX], [FeXI] and [FeXIV] is extraordinary
and exceeds 10$^{40}$ erg\,s$^{-1}$ in 2005. It
persists in [FeVII] for at least 3 years. Depending on the efficiency of
reprocessing UV--X-ray radiation into coronal lines, this implies an
intrinsic luminosity of at least $10^{42-43}$ erg\,s$^{-1}$. 
\item  At least three to four
different kinematical components are present in the optical spectrum,  
with FWHMs in the range $<$200 -- 2000 km\,s$^{-1}$.
\item  The narrow Balmer lines are triple-peaked with two remarkable unresolved 
horns which do not have a counterpart in any other
emission line; possibly the signature of collisionless shocks.  
\item   The $Spitzer$ SED 
shows a $\sim$10$\mu$ bump and a rise toward longer wavelengths, 
reminiscent of emission by warm silicate dust.
It shares some similarity with the average $Spitzer$ PG quasar SED. 
However, the NIR falls short of the extrapolated MIR SED
by one order of magnitude, unlike quasar spectra.
For few, if any, of such extreme transients as SDSSJ0952+2143, 
MIR spectroscopy was ever done. 
\item  The MIR luminosity is huge and amounts to at 
least $L_{\rm 10-20 \mu m} = 3.5 \times 10^{43}$ erg\,s$^{-1}$ 
at the time of the $Spitzer$ observation.   
The inferred black body radius of 0.5 pc implies that we 
see an IR light echo from pre-existing dust. 
\end{itemize}
 
Taken together, these data paint the picture of a very
energetic outburst of radiation 
in the inner 1.5 kpc of the galaxy SDSSJ0952+2143. 
The high-energy part of this flare was not observed directly,
but we did see the reprocessing of this radiation
into emission lines, and the low-energy continuum variability.  

There are two possible explanations for the radiation outburst: 
a tidal disruption/accretion event onto the SMBH
of a non-active or mildly active galaxy, or an extreme supernova
explosion.
Since only one single AGN previously showed high-amplitude 
variability in its iron coronal lines 
and no known non-active galaxy has shown this phenomenon,
a rigorous  
comparison with known object classes can only be done
for SNe.  Since there appears to be  nothing in the spectrum of SDSSJ0952+2143
which differs enough from a few known extreme SNe, we cannot exclude
an SN interpretation.
If an SN, SDSSJ0952+2143 is 
the most distant, and most X-ray and MIR luminous, X-ray detected SNIIn; and
the most luminous known in high-ionization iron coronal lines.  
An extreme accretion flare in a low-luminosity AGN or non-active
galaxy, especially by stellar tidal disruption, 
remain possibilities at present
and would produce potentially very similar emission-line spectra.
The large observed iron coronal line luminosities are unprecedented 
among SNe and exceed previous record holders by a factor 100.

\acknowledgments
We thank Lee Armus, Dirk Grupe, Stefan Immler, Gottfried Kanbach, Jean-Pierre Luminet,
Dieter Lutz, Ximena Mazzalay, 
Hagai Netzer, Nathan Smith, and Eckhard Sturm for very useful discussions,
Thomas Kr\"uhler for running the $GROND$ pipeline
data reduction, and our referee for his/her insightful comments.
Part of this research was carried out while
MD, SK, AR, and MS stayed at the Aspen Centre for Physics, and we thank the
Centre for its support and hospitality. 
We acknowledge use of the SDSS, $GALEX$, 2MASS, NVSS, $ROSAT$, $Swift$ BAT,
and $XMM-Newton$ data base
and of NED and ADS.   
We have acquired new data with ESO's $NTT$, the $Xinglong$ 
telescope, $GROND$, $Spitzer$ and $Chandra$.
We thank all instrument teams, and 
especially the SDSS team for making 
available to the whole astrophysics community
their outstanding data base on which the discovery of this galaxy was based.  
AR thanks Lee Armus for discussing the IRS data reduction. 
We acknowledge $Spitzer$ grant P483. 
HZ acknowledges support from the Alexander von Humboldt Foundation,
the Chinese Natural Science Foundation through CNSF-10533050, 10573015,
the CAS knowledge innovation project No. 1730812341, and the national
973 project (2007CB815403).
AG acknowledges support by the Israeli Science Foundation, an
EU Seventh Framework Programme Marie Curie IRG fellowship, and the
Benoziyo Center for Astrophysics,
a research grant from the Peter and Patricia Gruber Awards,
and the William Z. and Eda Bess Novick New Scientists Fund at the
Weizmann Institute.
DX acknowledges support from the Chinese National
Science Foundation (NSFC) under grant NSFC 10873017,
and from program 973 (2009CB824800).

 \clearpage

\begin{deluxetable}{llll}[hb]
\tabletypesize{\small}
\tablecaption{Sequence of events \& observations}
\tablewidth{0pt}
\tablehead{
\colhead{date} & \colhead{type of observation\tablenotemark{a}} & waveband or filter & \colhead{comments and source brightness} } \\
\startdata
1990 Nov. & RASS & 0.1$-$2.4 keV & no X-ray detection\\
          &      &             & $f_{\rm x} < 5\,10^{-13}$ erg cm$^{-2}$ s$^{-1}$  (unabsorbed flux, 0.2-2 keV) \\
1993 Dec. & NVSS & 1.4 GHz & no radio detection \\
          &      &             & $f_{\rm R} < 1$ mJy \\
1998-01-23 & 2MASS & J,H,K$_{\rm s}$ & IR detection of extended emission; likely pre-flare state \\
          &      &             & J=15.4, H=14.5, K$_{\rm s}$=14.4 mag \\
2002-05-07 & $XMM$-$Newton$ slew & 0.2$-$10 keV & no X-ray detection\\
          &      &             & $f_{\rm x} < 5.5\,10^{-12}$ erg cm$^{-2}$ s$^{-1}$  (unabsorbed flux, 0.2-10 keV) \\
2004-12-20 & SDSS photometry & u,g,r,i,z & peculiar SED and i,z high-point, highest observed state\\
          &      &             & u=18.34, g=17.7, r=17.11, i=16.65, z=16.22 mag (SDSS DR7) \\  
2005 Mar. -- & $Swift$ BAT survey & 15$-$55 keV & no X-ray detection \\
2008 Mar.    &      &             & $f_{\rm x} < 8\,10^{-12}$ erg cm$^{-2}$ s$^{-1}$  (unabsorbed flux, 15-55 keV) \\
2005-12-30 & SDSS spectroscopy & 3900--9100\AA & strong emission lines incl. iron; continuum fainter than during photometry \\
2006-03-02 & $GALEX$             & NUV, FUV & blue color  \\ 
          &      &             & NUV=20.2 mag, FUV=19.9 mag  \\ 
2007 Dec.      &              &         & source noticed by us, searching the SDSS data base \\
2007 Dec. 4-15  & $Xinglong$ spectroscopy & 4000--8000\AA & highest-ionization iron lines, HeII, and broad H$\alpha$ decreased in strength  \\
2008 Jan. 1 & $GROND$ photometry & g,r,i,z,J,H,K$_{\rm s}$ & confirms the galaxy(core) got fainter in the optical band   \\
2008 Feb. 4 & $Chandra$ & 0.1--10 keV & first detection of faint X-ray emission; implies high amplitude of variability  \\
          &      &             & $f_{\rm x} \simeq 6.3\,10^{-15}$ erg cm$^{-2}$ s$^{-1}$  (unabsorbed flux, 0.1-10 keV) \\
2008 Feb. 6 & $NTT$ spectroscopy & 3800--8600\AA & confirms fading lines in broader wavelength range  \\
2008 April  & $TripleSpec$ spectroscopy & NIR & NIR emission lines detected; analysis ongoing \\
~~~~~~~May  &                         &               \\
2008 June 5 & $Spitzer$ IRS & 9.9--19.6$\mu$m & search for emission lines; first MIR SED measurement \\
          &      &             & $f_{\rm IR} = 12$ mJy at 12.5$\mu$m \\ 
2008 Dec. 25 & $Xinglong$ spectroscopy & 4000--8000\AA & some iron lines fainter \\
2009  & $Chandra$ & 0.1--10 keV & X-ray monitoring; to be carried out 
\enddata
 \tablenotetext{a}{The following abbreviations have been used: RASS ($ROSAT$ all-sky survey; Voges et al. 1999),
NVSS (NRAO VLA Sky Survey; Condon et al. 1998), 2MASS (Two Micron All Sky Survey; Skrutskie et al. 2006), 
$GALEX$ (Galaxy Evolution Explorer; Martin et al. 2005), $GROND$ (Gamma-Ray Burst Optical/NIR Detector, Greiner et al. 2008),
and $NTT$ (New Technology Telescope).}     
\end{deluxetable}


\begin{deluxetable}{lcccccc} 
\tabletypesize{\small}
\tablecaption{Emission-line ratios relative to [OIII]5007, measured in the 2005 SDSS spectrum, and the 2008 $NTT$ spectrum.}
\tablewidth{0pt}
\tablehead{
\colhead{line} & {$\lambda$} & {$I/I_{\rm [OIII]}$\tablenotemark{a}} & {FWHM} & {$I/I_{\rm [OIII]}$} & {ratio\tablenotemark{b}} \\
               &            &   SDSS                                &  SDSS       &  ESO $NTT$        &  ESO/SDSS                          }
\startdata
~[FeVII] & 3586 & 0.44 & 230 & ... & ...  \\
~[OII]   & 3727 & 0.51 & 350 & 0.57 & 1.16 \\
~[FeVII] & 3759 & 0.74 & 515 & 0.47 & 0.67 \\
~[NeIII] & 3869 & 0.22 & 150 & 0.25 & 1.18 \\
~H$\gamma$ & 4340 & 0.24 & 380 & 0.13 & 0.57 \\ 
~[OIII] &  4363 & 0.23 & 710 & 0.27 & 1.20  \\
~HeII   & 4686 & 0.72 & 820 & 0.25 & 0.36  \\
~H$\beta_{\rm n}$ & 4861 & 0.33 & 210! & 0.30 & 0.95 \\
~H$\beta_{\rm horn, blue}$ &   & 0.09: & unresolved & $<$0.03: & $<$0.39 \\ 
~H$\beta_{\rm horn, red}$ &   & 0.23 & unresolved & 0.07: & 0.32 \\ 
~H$\beta_{\rm b}$ &      & 0.79 & 2100 & $<$0.58 & $<$0.76 \\ 
~[FeVII] ?    &4942 & 0.22 & 690  & 0.10 & 0.47 \\   
~[OIII] & 5007  & 1.0\tablenotemark{a} & 330  & 1.0 & 1.04 \\ 
~[FeVII] & 5158 & 0.14 & 300 & $<$0.03 & $<$0.25 \\ 
~[FeVI] & 5176  & 0.38 &  1180:  & ... & ... \\
~[FeXIV]\tablenotemark{c} & 5303  & 0.44 & 620 & 0.06: & 0.13 \\ 
~ ?    & 5621 & 0.35 & 820  & 0.08 & 0.24 \\   
~[FeVII] & 5722  & 0.56 & 450 & 0.25 & 0.46 & \\
~[FeVII] & 6087  & 0.67 & 320 & 0.50  & 0.76 & \\
~[FeX] & 6375 & 1.17 & 370: & 0.14: & 0.12 &  \\
~[OI] & 6300  & 0.12 & 210! & 0.05 & 0.41 & \\
~H$\alpha_{\rm n}$ & 6564  & 1.06 & 210! & 0.78 & 0.77 & \\ 
~H$\alpha_{\rm horn, blue}$ &   & 0.32 & unresolved & 0.06 & 0.2  \\ 
~H$\alpha_{\rm horn, red}$ &   & 0.50 & unresolved & 0.1 & 0.24  \\  
~H$\alpha_{\rm b}$ &      & 9.7  & 2100 & 3.29 & 0.35 \\ 
~[NII] & 6584 & 0.25 & 210! & 0.32 & 1.38  \\   
~[SII] & 6716 & 0.31 & 210 & 0.17 & 0.56  \\
~[SII] & 6731 & 0.24 & 210 & 0.15 & 0.65 \\
~[FeXI] & 7892 & 0.87 & 440 & 0.10 & 0.12       
\enddata
\tablenotetext{a}{Line ratios reported in this table are based on single Gaussian fits to the emission lines,
except for the broad components of H$\alpha$ and H$\beta$ which were represented by two Gaussians
(see text for details).
For each epoch, line ratios have been normalized to [OIII]$\lambda$5007. The
observed [OIII] flux  
is
$f_{\rm [OIII],SDSS}$ = 1.8\,10$^{-15}$ erg\,cm$^{-2}$\,s$^{-1}$
and $f_{\rm [OIII],ESO}$ = 1.9\,10$^{-15}$ erg\,cm$^{-2}$\,s$^{-1}$,
in the 2005 SDSS and 2008 $NTT$ spectrum, respectively.} 
\tablenotetext{b}{The ratio is the direct flux ratio between 2005 SDSS and 2008 $NTT$ emission lines.} 
\tablenotetext{c}{Potentially blended with [CaV]5309 in the ESO spectrum. \\
Line ratios marked with a colon are uncertain measurements, and an exclamation mark indicates that this
parameter was fixed in the fitting 
procedure. A question mark indicates an uncertain line identification. }
\end{deluxetable}


\begin{deluxetable}{lll} 
\tabletypesize{\small}
\tablecaption{Summary of some key parameters}
\tablewidth{0pt}
\tablehead{
\colhead{parameter} & {measurement} & {comment} }
\startdata
X-ray luminosity & 1\,10$^{41}$ erg\,s$^{-1}$ & 2008 $Chandra$ (0.1-10 keV) \\
R magnitude   & --20.8 & 2004 SDSS photometry \\
MIR luminosity  &  3.5\,10$^{43}$ erg\,s$^{-1}$  & 2008 $Spitzer$ (10 -- 20 $\mu$m) \\ 
broad H$\alpha$ luminosity & 3 10$^{41}$ erg\,s$^{-1}$ & 2005 SDSS spectrum \\
~[FeX]6375 luminosity & 3.5 $10^{40}$ erg\,s$^{-1}$  & brightest detected coronal line in 2005 \\
~[FeVII]6087 luminosity & 1.6 $10^{40}$ erg\,s$^{-1}$  & brightest detected coronal line in 2008 \\
Balmer decrement H$\alpha_{\rm b}$/H$\beta_{\rm b}$  & 12 & 2005 SDSS spectrum \\
SMBH mass & $7\,10^6 M_{\odot}$ & from stellar absorption features of the host galaxy; paper I 
\enddata
\end{deluxetable}

 \clearpage

\begin{figure*}[b]
\plotone{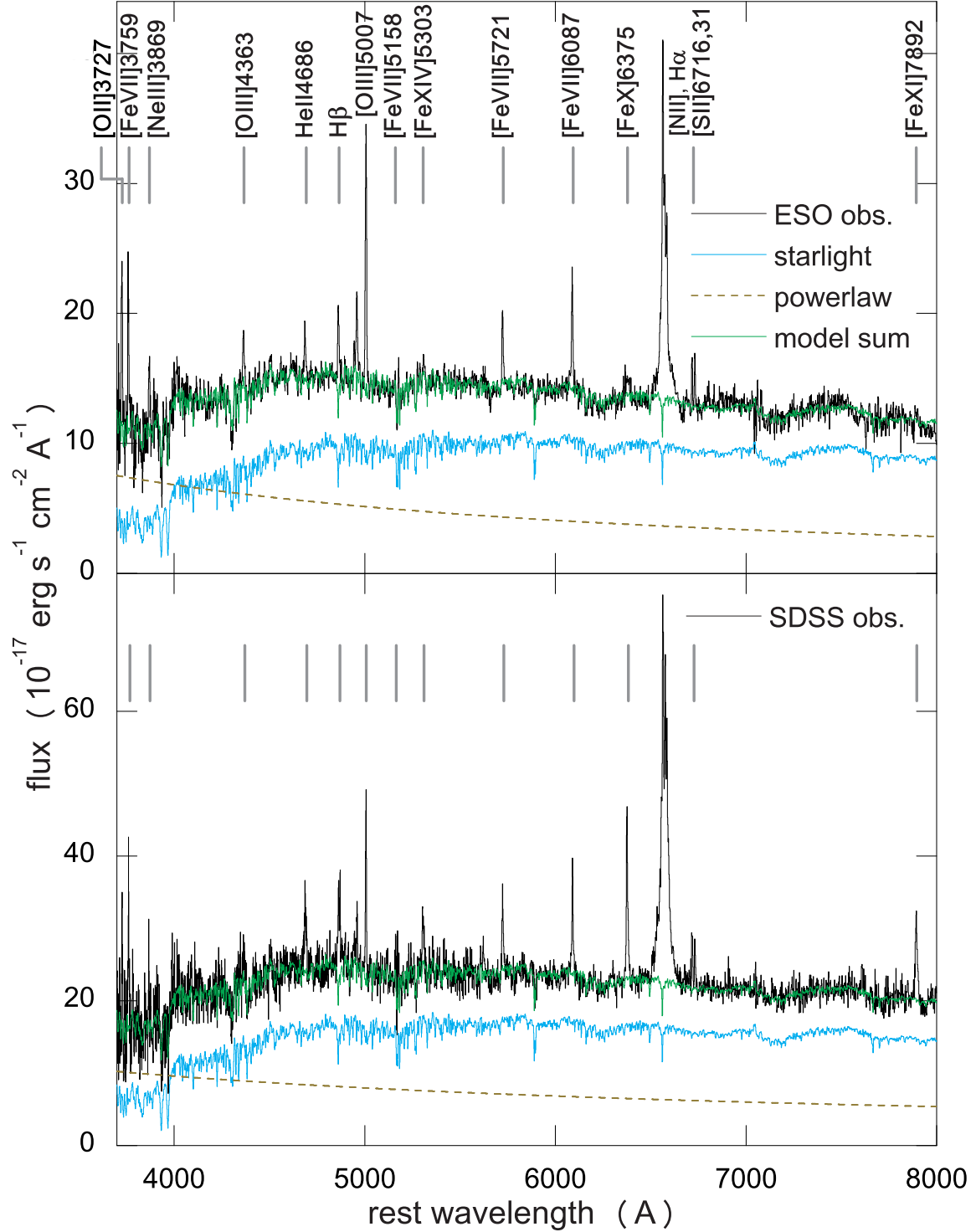}
\caption{2005 SDSS spectrum (lower panel) and the 2008 ESO $NTT$ spectrum (upper panel) of SDSSJ0952+2143. 
Emission lines are labeled. The continuum emission was decomposed into stellar (blue solid line) and powerlaw
[index $\alpha_{\lambda}=-1.2$ (SDSS) and $-0.7$ (NTT); light green dashed line] component as given in the graph. }
\end{figure*}

\begin{figure*}
\plotone{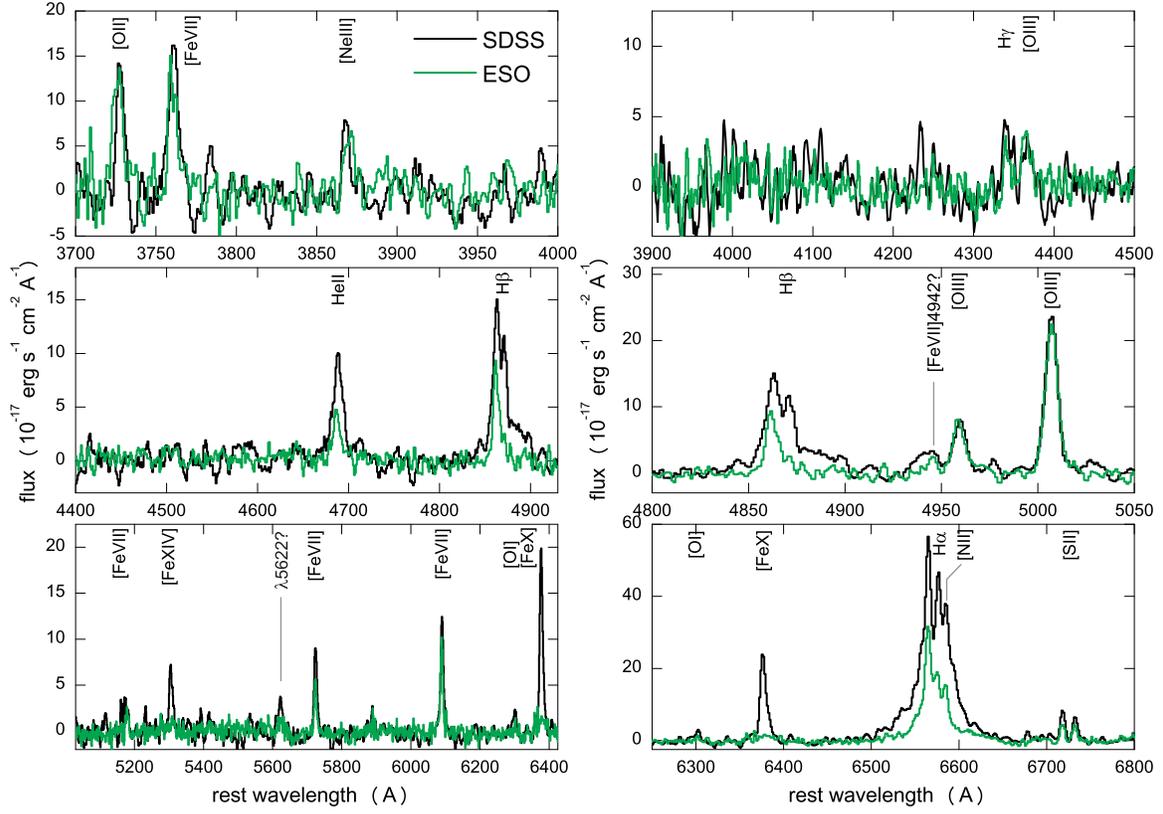}
\caption{Zoom on individual emission-line complexes after continuum subtraction
(note the different flux and wavelength scale in each
panel, optimized to display emission lines of different strengths). 
SDSS (black) and ESO $NTT$ (green) spectrum are overplotted
on top of each other. The resolution of the SDSS spectrum was
degraded to match that of the $NTT$ spectrum.  
Prominent emission lines are marked. } 
\end{figure*}

\begin{figure*}
\plotone{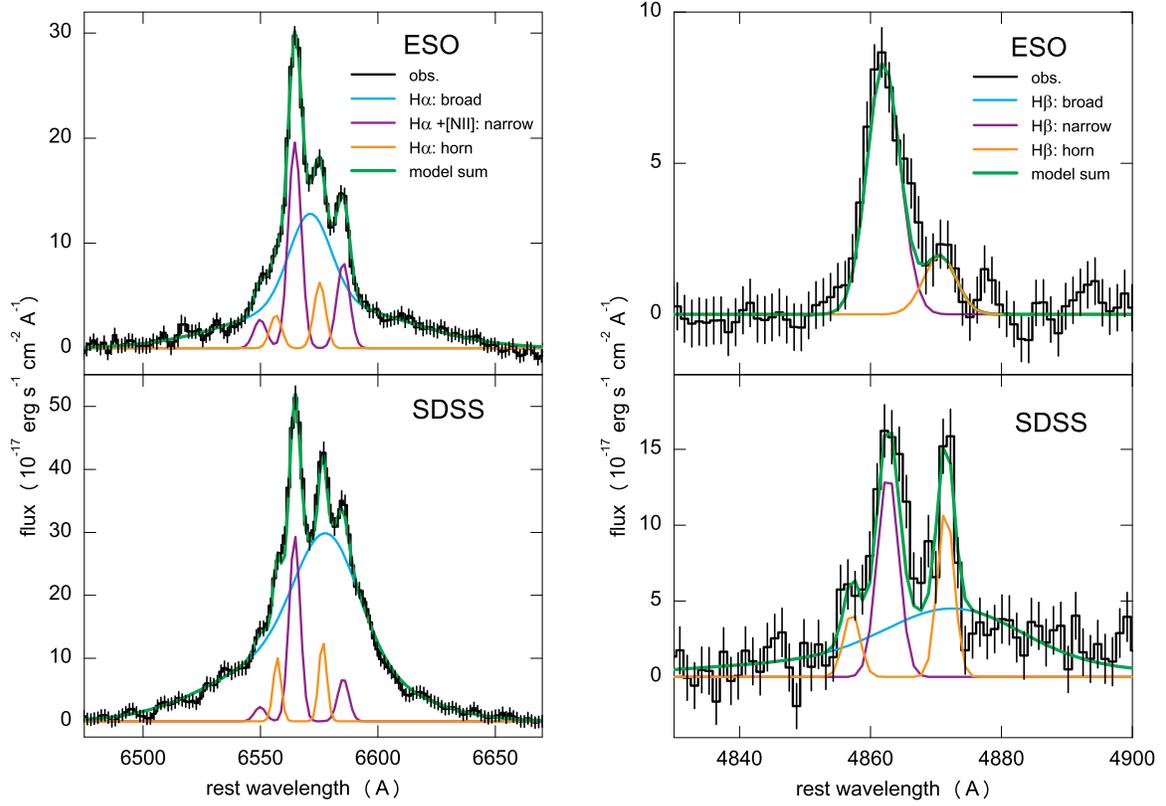}
\caption{Decomposition of the Balmer-line profiles of H$\alpha$ (left) and H$\beta$ (right) into several components.
The lower panel shows the 2005 SDSS data, the upper panel the 2008 ESO $NTT$ data (note that the
flux scale is different on each axis). The different components
of the profile are labeled in each figure. Note the change in the broad component and the two narrow
horns which are all fainter in the 2008 spectrum.   }
\end{figure*}

\begin{figure*}
\plotone{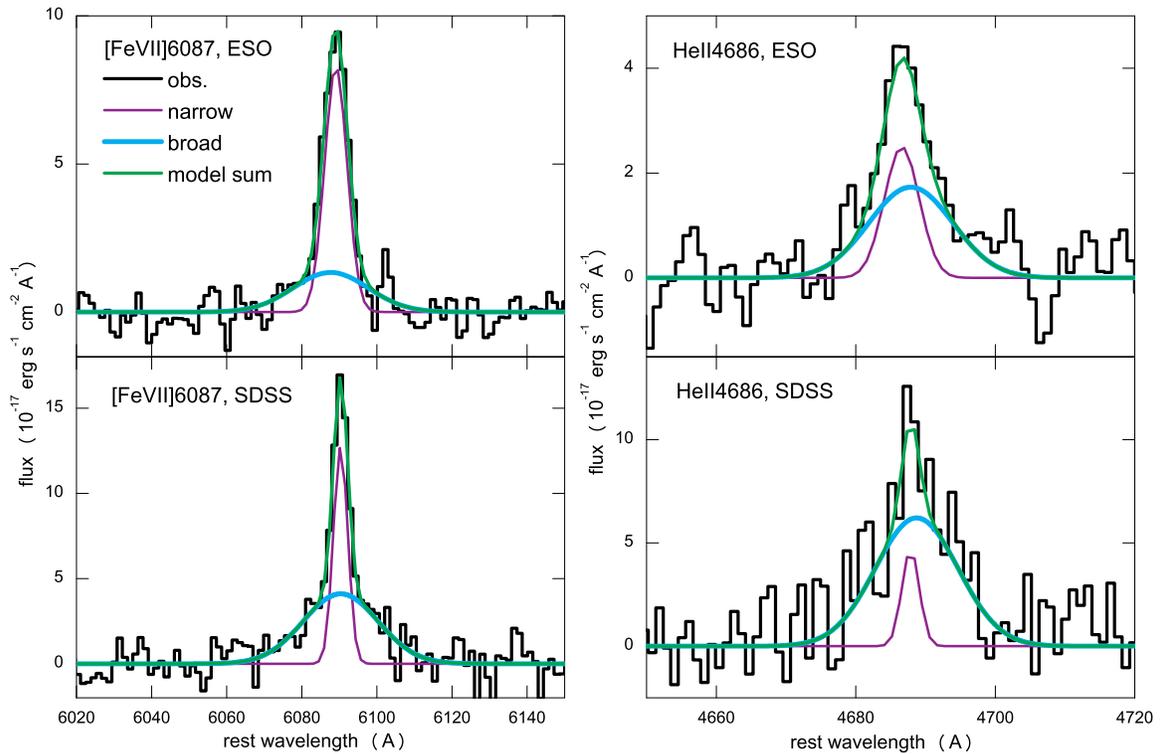}
\caption{Decomposition of the two bright high-ionization lines HeII4686 and [FeVII]6087 into two components. 
The broad base which is present in the 2005 SDSS spectrum is much fainter in the 2008 ESO $NTT$ spectrum.  
[{\it{See the electronic edition of the journal for a colour version of this figure.}}]   }
\end{figure*}

\begin{figure*}
\plottwo{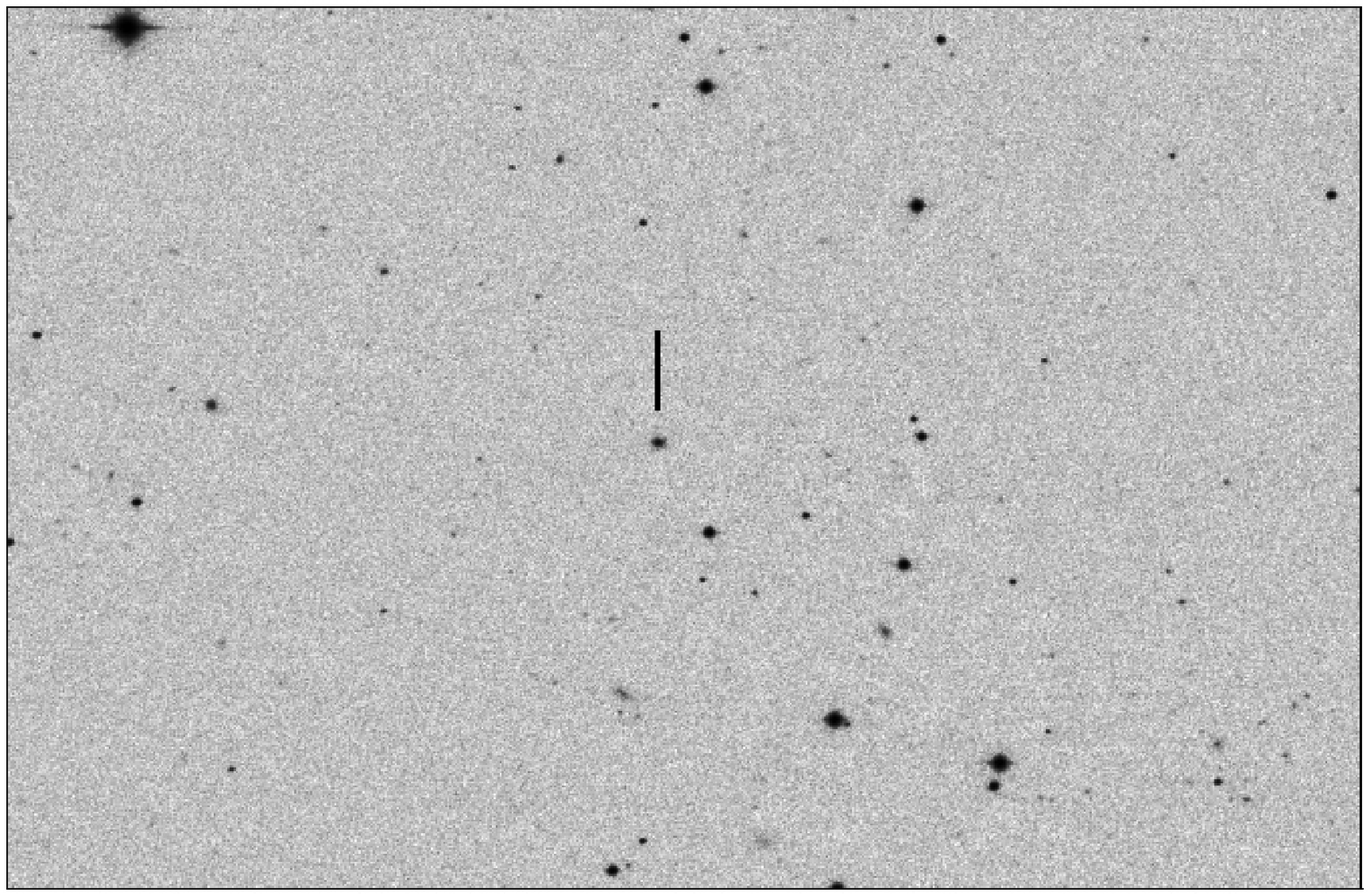}{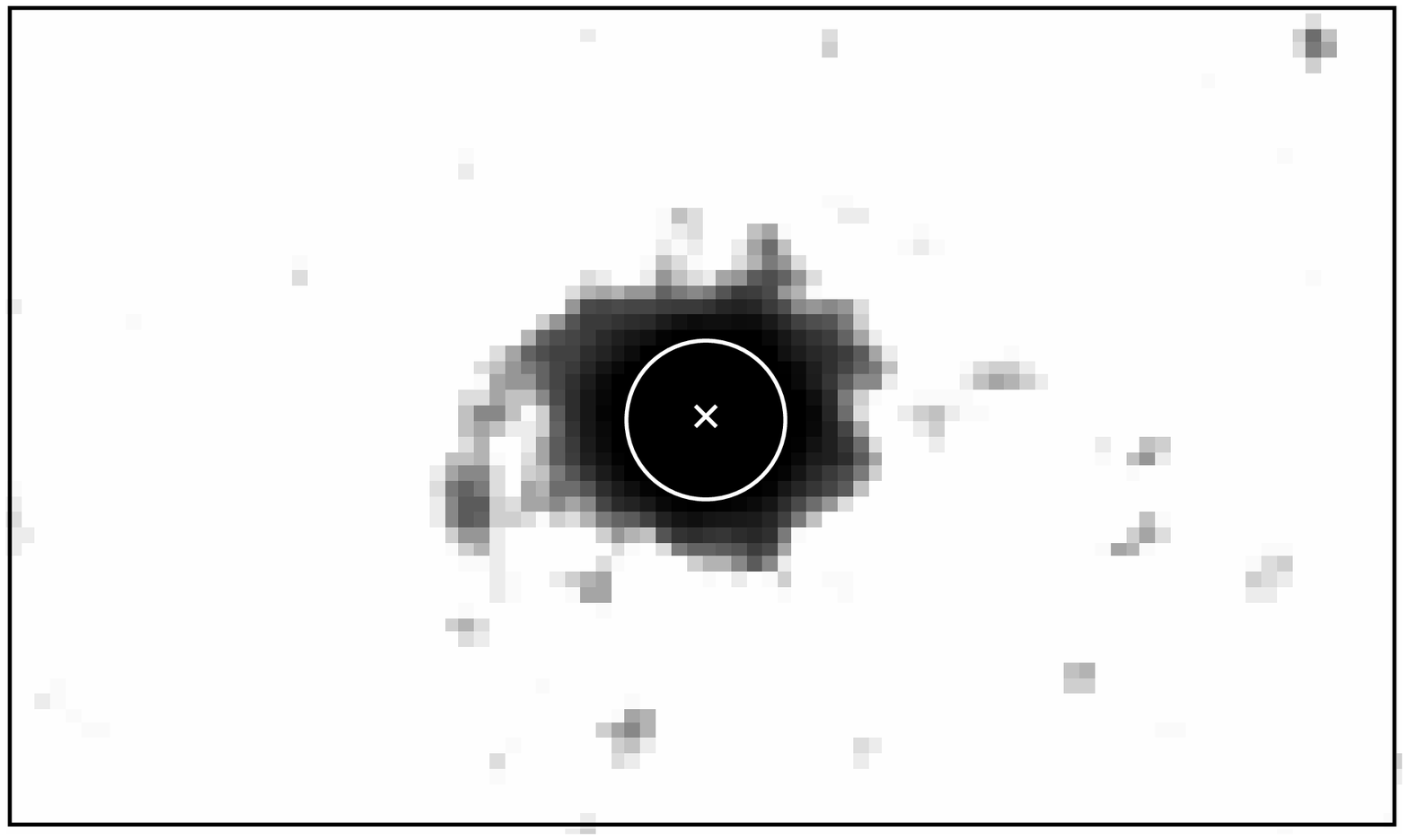} 
\caption{{\em Left}: $GROND$ J-band image of the field of SDSSJ0952+2143 which
is marked with the black bar. 
{\em Right}: Zoom on the K$_{\rm s}$-band image of SDSSJ0952+2143 and overlay of
the $Chandra$ X-ray position (white cross) plus error circle of 1$^{\prime\prime}$ radius.    }
\end{figure*}

\begin{figure*}
\plotone{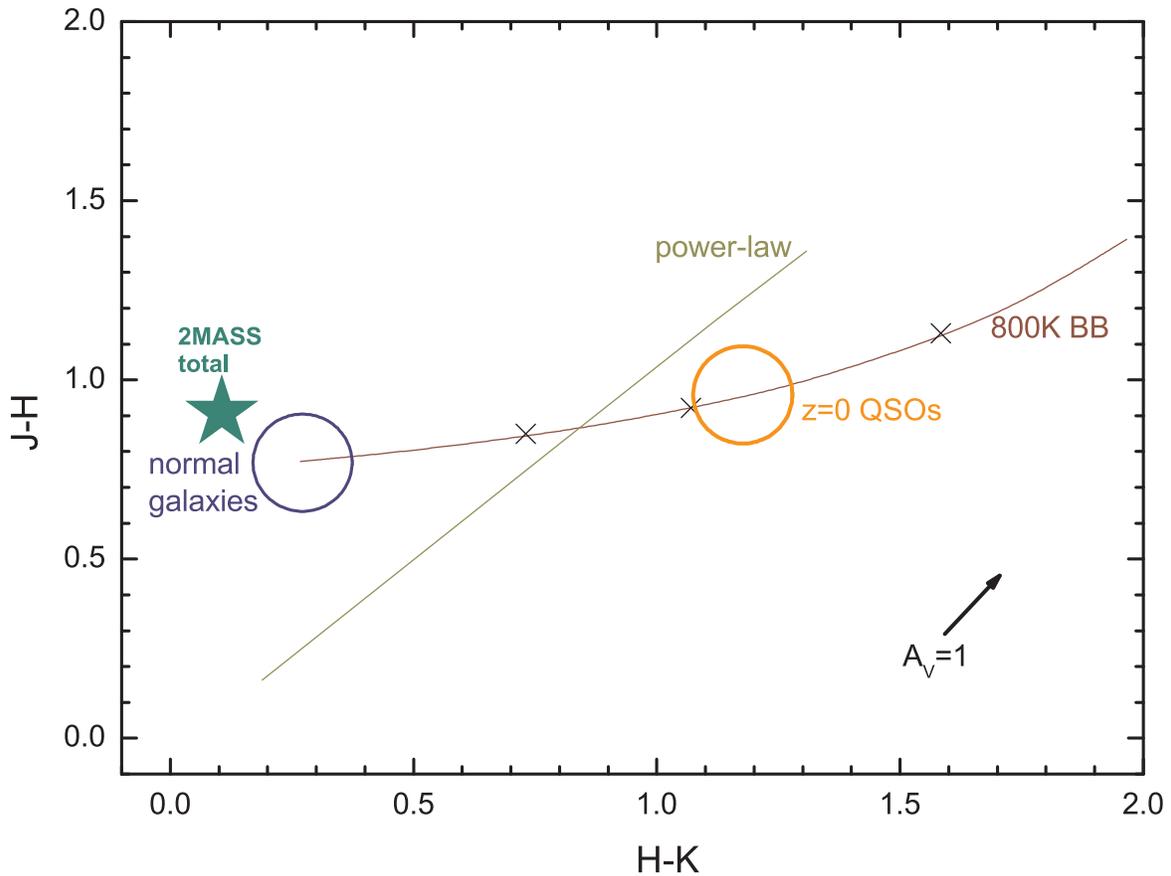} 
\caption{Locus of the 2MASS NIR colours of SDSSJ0952+2143 in comparison to
the loci of quiescent galaxies and low-redshift QSOs (Hyland \& Allen 1982). 
For comparison, the two curves represent a mix of a normal stellar
population and a powerlaw or black body (800 K) emission component with
increasing strength from left to right. 
The contribution of the stellar population decreases from left to right
along the black body curve. 
Reddening by $A_V=1$ is indicated by the arrow. The 2MASS data point 
lies in a region where normal galaxies are found. 
[{\it{See the electronic edition of the journal for a colour version of this figure.}}] 
 }
\end{figure*}

\begin{figure}[htbp]
\begin{center}
\plotone{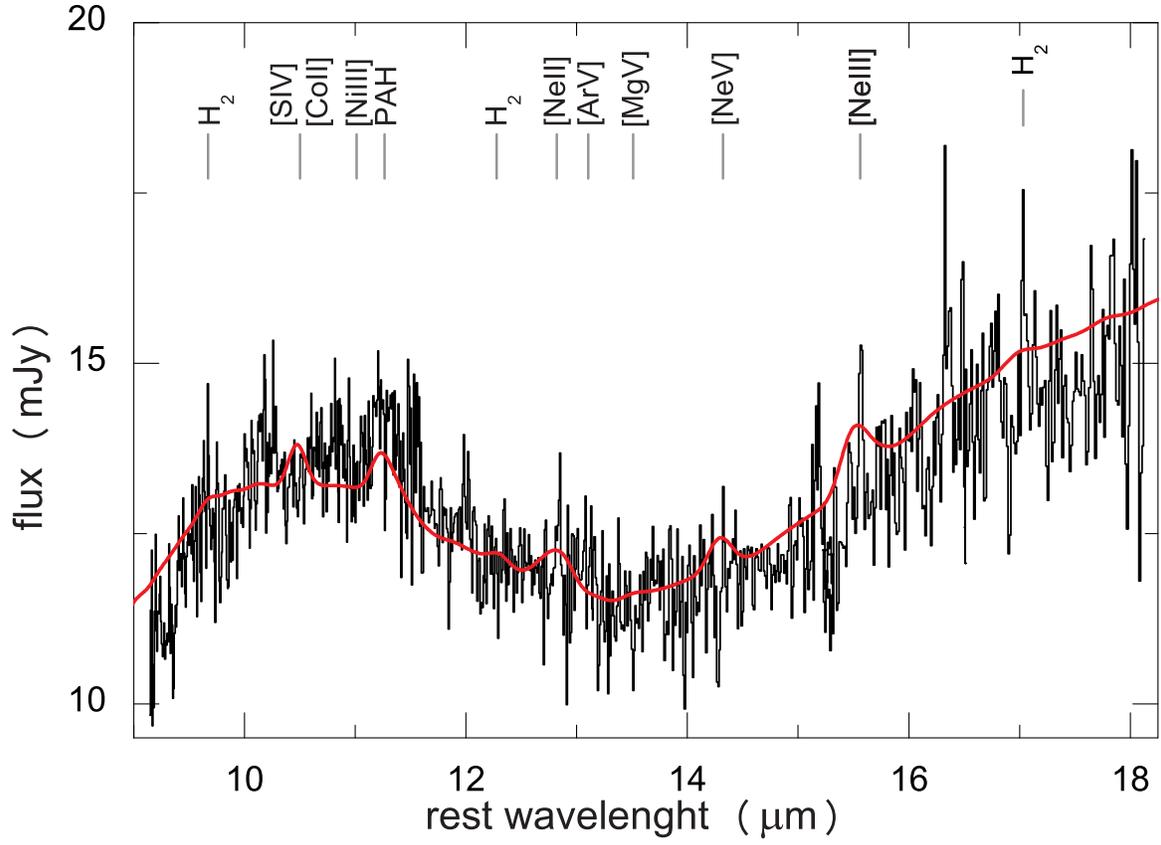} 
\caption{$Spitzer$ Short-High IRS spectrum of SDSSJ0952+2143 taken in 2008 June. 
The broad bump
around 11\,$\mu$m and the increase towards longer wavelengths are 
reminiscent of dust emission from silicates. The red line overplots
the $Spitzer$ mean quasar SED of Netzer et al. (2007) which represents the 
shape of the overall spectrum
well (its normalization was chosen such that it matches the flux of SDSSJ0952+2143
at 14$\mu$m). 
The wavelengths of several emission lines typically observed in starburst galaxies, AGN, and SNe
are marked.   
 }
\label{fig:irs}
\end{center}
\end{figure}

\begin{figure*}
\plotone{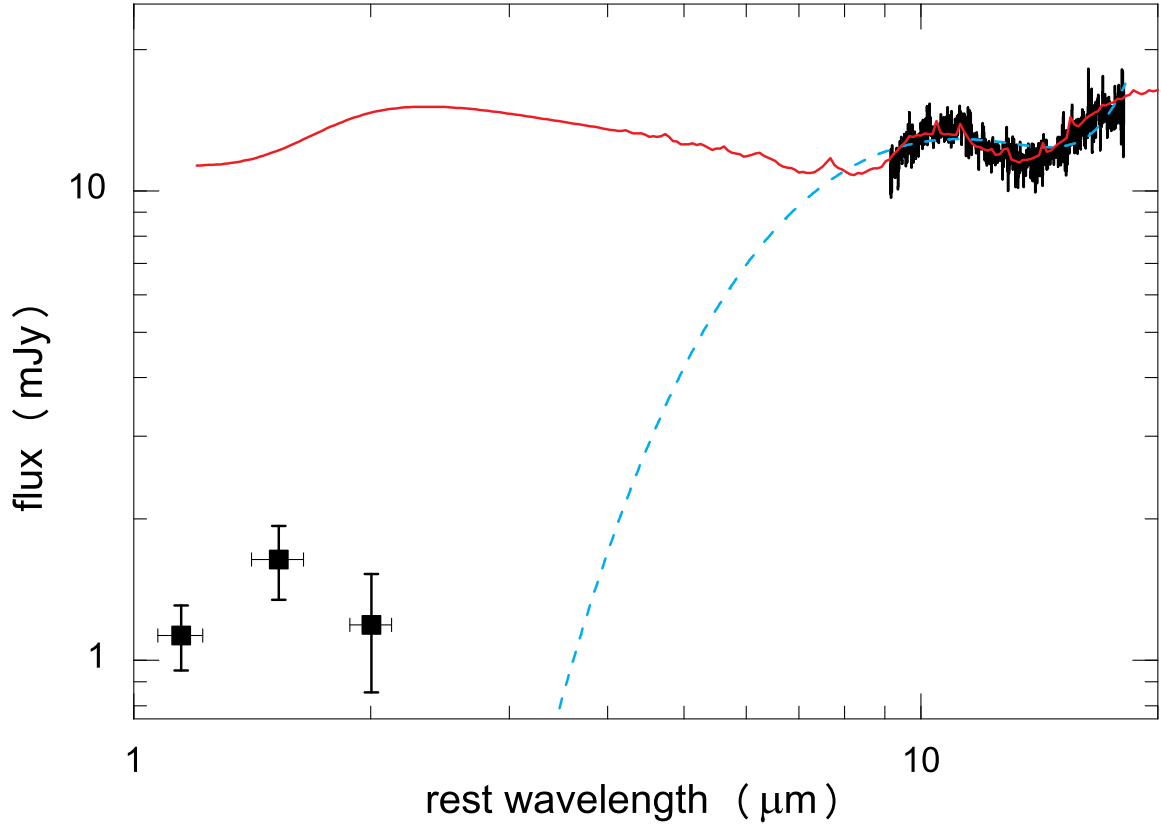}
\caption{IR SED of SDSSJ0952+2143 observed with $Spitzer$ and 2MASS (the latter
consistent with $GROND$ within an error of $\sim$0.5 mag). Overplotted is the
mean $Spitzer$ quasar SED (red; Netzer et al. 2007) and a simple two-component black body fit to
the $Spitzer$ data (blue). 
}
\end{figure*}

\end{document}